\documentclass[useAMS,usenatbib]{mn2e}

\usepackage{graphicx}

\title[Does Sombrero have a (classical) bulge?]{Surprises in Image Decomposition of Edge-on Galaxies: does Sombrero have a (classical) bulge?}
\author[Gadotti \& S\'anchez-Janssen]{Dimitri A. Gadotti\thanks{E-mail: dgadotti@eso.org} and 
Rub\'en S\'anchez-Janssen\\
European Southern Observatory, Casilla 19001, Santiago 19, Chile}
\begin{document}


\pagerange{\pageref{firstpage}--\pageref{lastpage}} \pubyear{2011}

\maketitle

\label{firstpage}

\begin{abstract}
The spheroid of the Sombrero galaxy, NGC 4594, is considered a prototype of classical, merger-built bulges. We use a Spitzer, IRAC 3.6$\mu m$ image to perform a detailed structural analysis of this galaxy. If one fits to this image only bulge and disc components, the bulge occupies a locus in the mass--size relation close to that of elliptical galaxies. When an outer stellar spheroid is added to improve the fit, the bulge S\'ersic index drops by a factor of $\approx2$, and, if taken at face value, could mean that this bulge is actually a disc-like, pseudo-bulge, or a bar viewed end-on. The bulge effective radius and the bulge-to-total ratio also drop dramatically, putting the bulge in a position closer to that of bulges in the mass--size relation. We discuss implications from these findings, including the locus of the Sombrero bulge in the black hole mass vs. bulge mass relation. With this new bulge mass estimate, current dynamical estimates for the mass of the central black hole in Sombrero are more than 10 times larger than expected, if only the bulge mass is considered. A better agreement is found if the sum of bulge and outer spheroid masses is considered. Furthermore, residual images show the presence of a stellar ring and a stellar, inner ring or disc, with unprecedented clarity. We also show that Sombrero is an outlier in scaling relations of disc galaxies involving the disc, the spheroid and the globular cluster system, but not so when its structural components are considered independently. In this context, the globular cluster system of Sombrero might not be representative of disc galaxies. Finally, we discuss the possibility that Sombrero formed as an elliptical galaxy but accreted a massive disc, which itself has secularly evolved, resulting in a complex and peculiar system.
\end{abstract}


\section{Introduction}
\label{sec:intro}

The Sombrero galaxy (NGC 4594, or M 104) is a local, massive disc galaxy. Classified as an unbarred Sa \citep{deVdeVCor91}, it is at a distance of 9.1Mpc \citep[][from infrared surface brightness fluctuations]{JenTonBar03}, and has a stellar mass of $\sim2.3\times10^{11}$M$_\odot$ \citep[e.g.][]{temten06}. At this distance, 1" corresponds to about 44pc. It is seen almost perfectly edge-on, with an inclination angle of $\approx84^\circ$ \citep{emsbacmon96}. It exhibits a remarkable spheroidal structure which extends to distances much larger than usually seen in other inclined disc galaxies.

\begin{figure}
   \centering
   \includegraphics[keepaspectratio=true,width=7cm,clip=true]{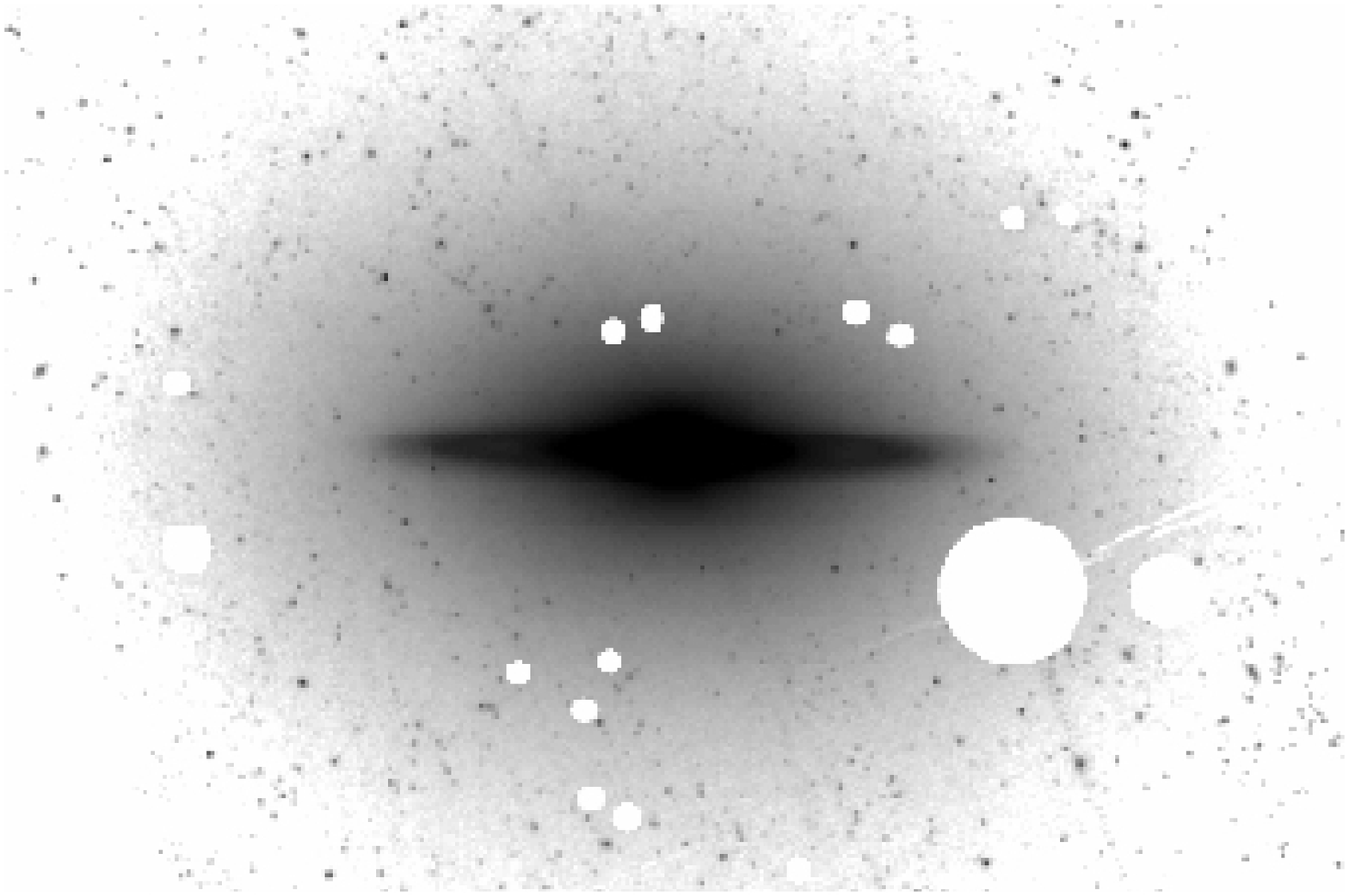}
   \includegraphics[keepaspectratio=true,width=7cm,clip=true]{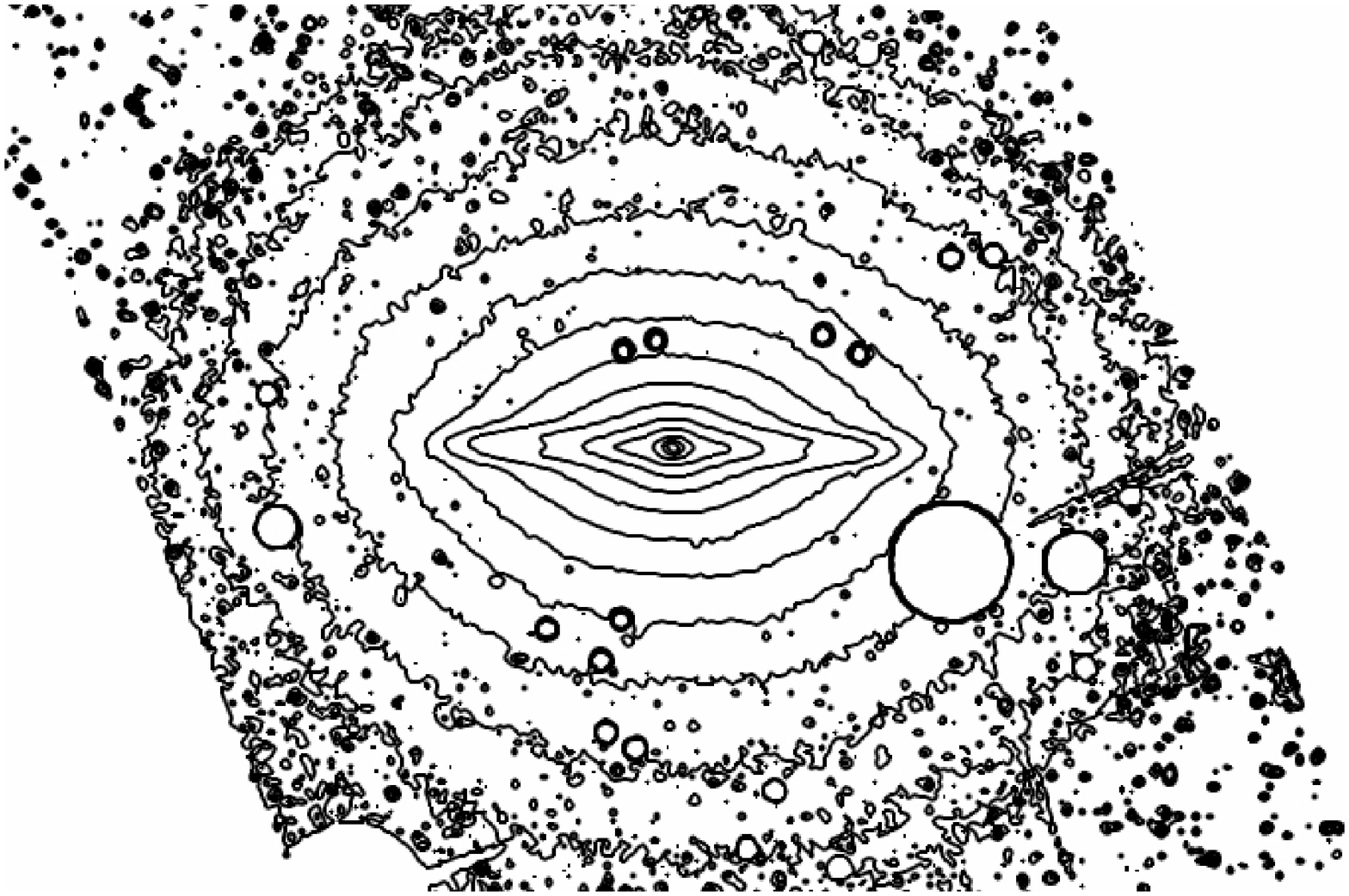}
   \caption{Top: IRAC channel 1 image of the Sombrero galaxy (NGC 4594) at 3.6$\mu m$, used in this study. Bottom: isophotal contours from the same image. Foreground stars are removed. The stellar halo stands out clearly and appears rounder than the central bulge.}
   \label{fig:irac}
\end{figure}

Figure \ref{fig:irac} shows an archive Spitzer IRAC image of this galaxy at 3.6$\mu m$ \citep[from][]{kenarmben03}, where there is very little dust absorption or emission, as well as little contamination from hot, young stars \citep[see e.g.][]{shereghin10}. It is thus a superb source to study the bulk structural properties of the galaxy. This deep image shows that the extended stellar halo seems rounder than the more compact central bulge, closer to the disc plane. In early studies based on shallower or optical images the full extent of the halo is not evident, and thus Sombrero has been traditionally considered a bulge+disc system only, with a very high bulge-to-total ratio \citep[$B/T=0.86$ in][see also \citealt{larbrohuc01,dramcgdop03,GulRicGeb09}]{ken88}. This difference between the shape of the extended halo and the central bulge might suggest different formation processes for both components. Alternatively, if halo and bulge form as a single entity, then they could have followed different evolutionary paths. A further alternative is that the variation in ellipticity is a projection effect \citep[see][]{Ryd91} on a single spheroidal component.

Since most previous works did not distinguish the halo from the bulge, the whole spheroid was well fitted by a S\'ersic law with high S\'ersic index, i.e. $n\sim4$, although then the disc had to be fitted using complicated non-standard functions \citep[see][]{ken88}. The spheroid is obviously rounder and vertically more extended than the disc, and it also has no conspicuous star formation. For these reasons, the spheroid in NGC 4594 has been considered a prototype of classical bulges \citep[e.g. in][]{KorKen04}, i.e. those formed through violent processes, as opposed to disc-like bulges, which are thought to form from disc material brought to the centre via non-axyssimetrical components (such as bars), which form as a result of disc instabilities \cite[see][]{Ath05b}. In fact, it has been advocated by e.g. \citet{FisDro08} that the bulge S\'ersic index can be used to separate classical bulges from disc-like bulges, with a threshold at $n\approx2$. In \citet{Gad09b}, it is argued that, although a very useful parameter, $n$ has to be used with care when distinguishing the different bulge categories, since uncertainties in its measurement can be relatively high. With a different methodology to separate disc-like bulges, it is shown that the median value of $n$ for disc-like bulges is 1.5, with a standard deviation of 0.9. For classical bulges, one finds that the median $n=3.4$, with standard deviation of 1.3. Furthermore, disc-like bulges have typical values of $B/T$ below 0.1, whereas classical bulges have a distribution of $B/T$ that peaks at $\approx0.4$.

In this study, we perform a detailed structural analysis of the Sombrero galaxy, using the Spitzer image shown in Fig. \ref{fig:irac}, in Sect. \ref{sec:struc}. We show that if an outer stellar spheroid, or halo, is included, both the bulge $n$ and $B/T$ drop dramatically. We discuss the implications of this finding in Sect. \ref{sec:disconc}.

\section{Structural Analysis}
\label{sec:struc}

\subsection{Fits to spheroid and disc}

We used {\sc budda} to decompose NGC 4594 into its different structural components. Firstly, we fitted a model with bulge and disc components only (model {\sc bd}). The code uses the entire galaxy image and each component is described as a series of concentric generalised ellipses. The bulge follows a S\'ersic profile and 5 parameters are left free to fit it: the S\'ersic index, effective radius, effective surface brightness, position angle and ellipticity. The disc is fitted as an edge-on disc, following Eq. (7) in \citet{vanSea81}, and 3 parameters are fitted: scale-length, scale-height and central surface brightness. Although the disc is not perfectly edge-on, tests showed that an edge-on disc model produces marginally better results in this case. The reader is referred to \citet{deSGaddos04} and \citet{Gad08} for details on the workings of the {\sc budda} code.

Two points deserve further discussion here. Firstly, the Sombrero galaxy has a weak AGN, more specifically a LINER \citep{benbucdal06}. These authors produced 2D fits to images of Sombrero at wavelengths ranging from 3.6$\mu$m (using the same image as in here) to 850$\mu$m, and found that this nucleus is brighter at 24$\mu$m. At 3.6$\mu$m, due to problems in modeling the PSF, they excluded from the fit the inner 10'', thus excluding the AGN component. \citet{jargebshe11} recently performed a bulge/disc decomposition of Sombrero and avoided the AGN by excluding the inner 0.17'' from their fit. A discussion on how the structural analysis presented here compares with these studies is given further below. Typically, only bright type 1 AGN should be modeled in fits such as the one presented here \citep[see][]{Gad08}, and the effect of not including an AGN component in these cases is usually to overestimate the bulge-to-total ratio, as well as the bulge S\'ersic index. In order to check whether an AGN component should be included in the fits to the Spitzer 3.6$\mu$m image, we produced a test fit with such component. As expected, since the AGN is faint, adding an AGN component yields very similar results as when the component is absent, and thus its inclusion is not relevant. The most substantial change is in the bulge S\'ersic index, which is $\approx10\%$ lower when the AGN component is added. Such a difference is however within the 1$\sigma$ uncertainty, and the fit produced is slightly worse. In this context, it is important to note that the spatial resolution of the Spitzer image is relatively low, with a pixel size of 0.75'' and a PSF FWHM of $\approx$2''. In \citet{Gad08}, it is shown that at low spatial resolution even the light from bright type 1 AGN can be smeared out, and an AGN component should not be included in this case. A further reason to not include an AGN component in the fits here is that it can artificially reinforce some of the results presented below, namely that the S\'ersic index of the Sombrero bulge can be lower than expected and that a fit with bulge and disc only is not a good fit. Thus, the fits presented here do not include an AGN component.

Secondly, a good modeling of the PSF is critical for this study, since if it is not well modeled the resulting bulge parameters from the fits can be wrong. The PSF is modeled as a circular Moffat function \citep[see][]{TruAguCep01}. The FHWM and $\beta$ Moffat parameters were estimated from fits to several point sources in the field of the Spitzer image, using standard routines in {\sc iraf}\footnote{{\sc iraf} is distributed by the National Optical Astronomy Observatories,
which are operated by the Association of Universities for Research in Astronomy, Inc., under
cooperative agreement with the National Science Foundation.}. We have also produced test fits using the PSF image built for the Spitzer Survey of Stellar Structure in Galaxies \citep[S$^4$G -- ][]{shereghin10} by combining stars in many Spitzer 3.6$\mu$m images, which returned similar results.

\begin{figure}
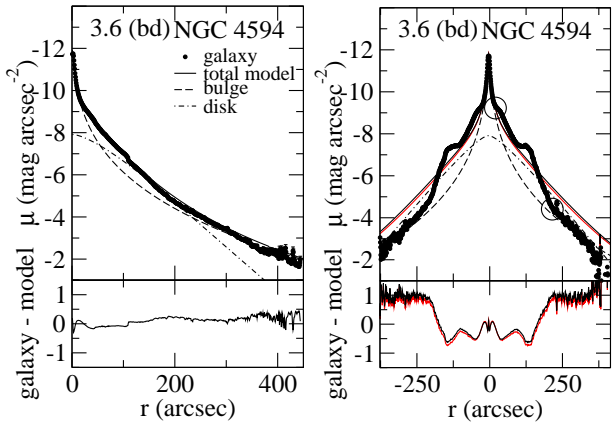

   \centering
   \includegraphics[keepaspectratio=true,width=4cm,clip=true]{N4594_bd.eps}
   \includegraphics[keepaspectratio=true,width=4cm,clip=true]{N4594vec_bd_3.eps}
   \caption{Left: surface brightness radial profiles (in arbitrary units) obtained through ellipse fits to the IRAC image and to the {\sc budda} model images of a decomposition including bulge and disc only. Right: the same radial profiles, but obtained through a cut along the disc major axis. Lower panels show residual profiles after subtracting the full {\sc budda} model profile from the galaxy profile. The two circles mark the positions of the breaks in the galaxy radial surface brightness profile discussed in the text. The red solid lines have the same meaning as for the solid black lines, but when the fit is done masking out the outer ring only.}
   \label{fig:profs_bd}
\end{figure}

\begin{figure}
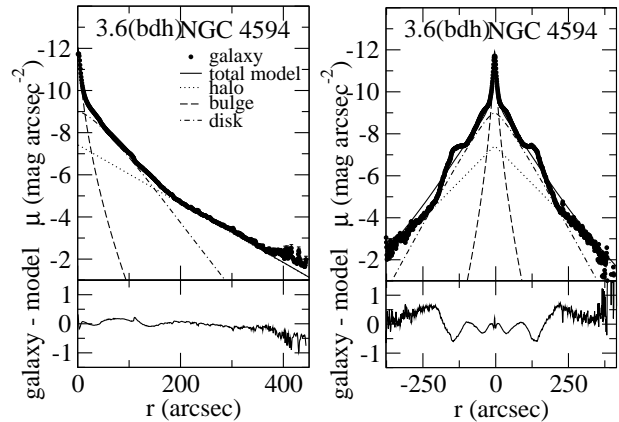

   \centering
   \includegraphics[keepaspectratio=true,width=4cm,clip=true]{N4594_bdh.eps}
   \includegraphics[keepaspectratio=true,width=4cm,clip=true]{N4594vec_bdh_2.eps}
   \caption{Same as Fig. \ref{fig:profs_bd} but for a {\sc budda} decomposition including bulge, disc and an exponential halo.}
   \label{fig:profs_bdh}
\end{figure}

\begin{figure}
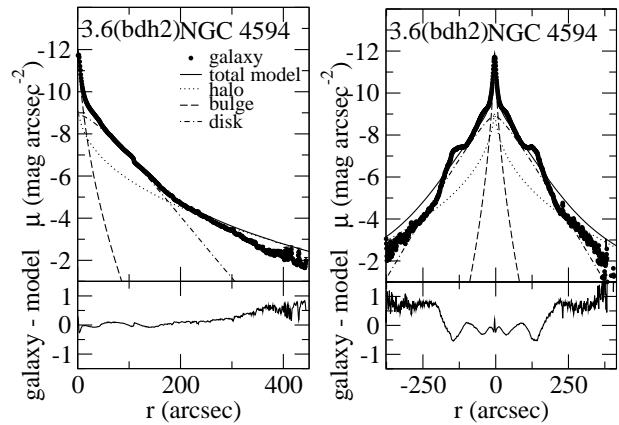

   \centering
   \includegraphics[keepaspectratio=true,width=4cm,clip=true]{N4594_bdh2.eps}
   \includegraphics[keepaspectratio=true,width=4cm,clip=true]{N4594vec_bdh2_2.eps}
   \caption{Same as Figs. \ref{fig:profs_bd} and \ref {fig:profs_bdh} but for a {\sc budda} decomposition including bulge, disc and a S\'ersic halo with S\'ersic index constrained to be larger than 2.}
   \label{fig:profs_bdh2}
\end{figure}

\begin{figure}
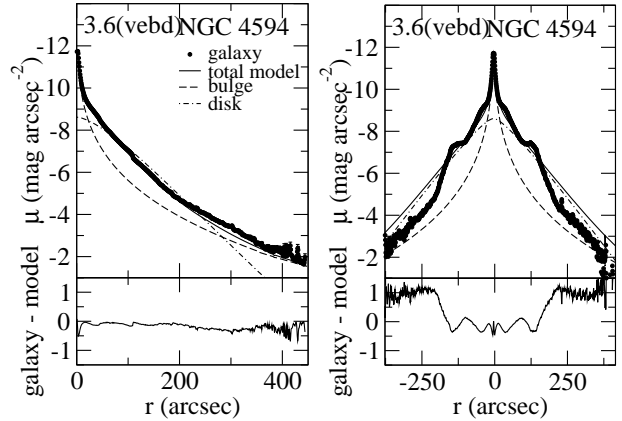

   \centering
   \includegraphics[keepaspectratio=true,width=4cm,clip=true]{N4594_veb.eps}
   \includegraphics[keepaspectratio=true,width=4cm,clip=true]{N4594vec_veb.eps}
   \caption{Same as Figs. \ref{fig:profs_bd} to \ref{fig:profs_bdh2} but for a {\sc budda} decomposition including a bulge, {\em with ellipticity varying with radius}, and a disc, but with no halo.}
   \label{fig:profs_veb}
\end{figure}

\begin{figure}
   \centering
   \includegraphics[keepaspectratio=true,width=6cm,clip=true]{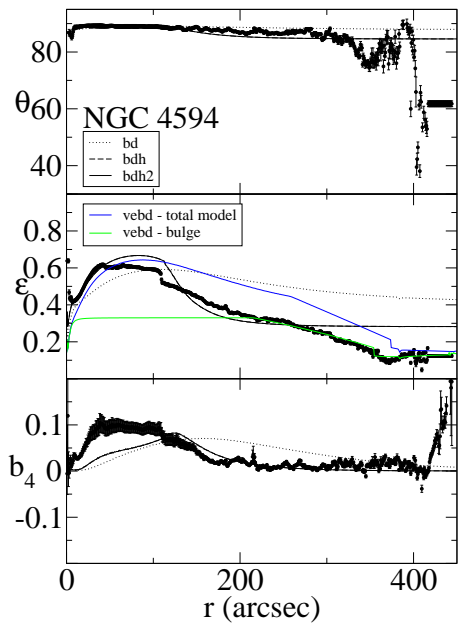}
   \caption{Radial profiles of position angle (top), ellipticity (centre) and Fourier b$_4$ coefficient obtained through ellipse fits to the IRAC image (points with error bars) and to the 3 full {\sc budda} models, as indicated. There is practically no difference between models {\sc bdh} and {\sc bdh2} concerning these profiles. The middle panel also shows the ellipticity profiles of the bulge component only and of the total model in model {\sc vebd}.}
   \label{fig:geo}
\end{figure}

\begin{figure*}
   \centering
   \includegraphics[keepaspectratio=true,width=5cm,clip=true]{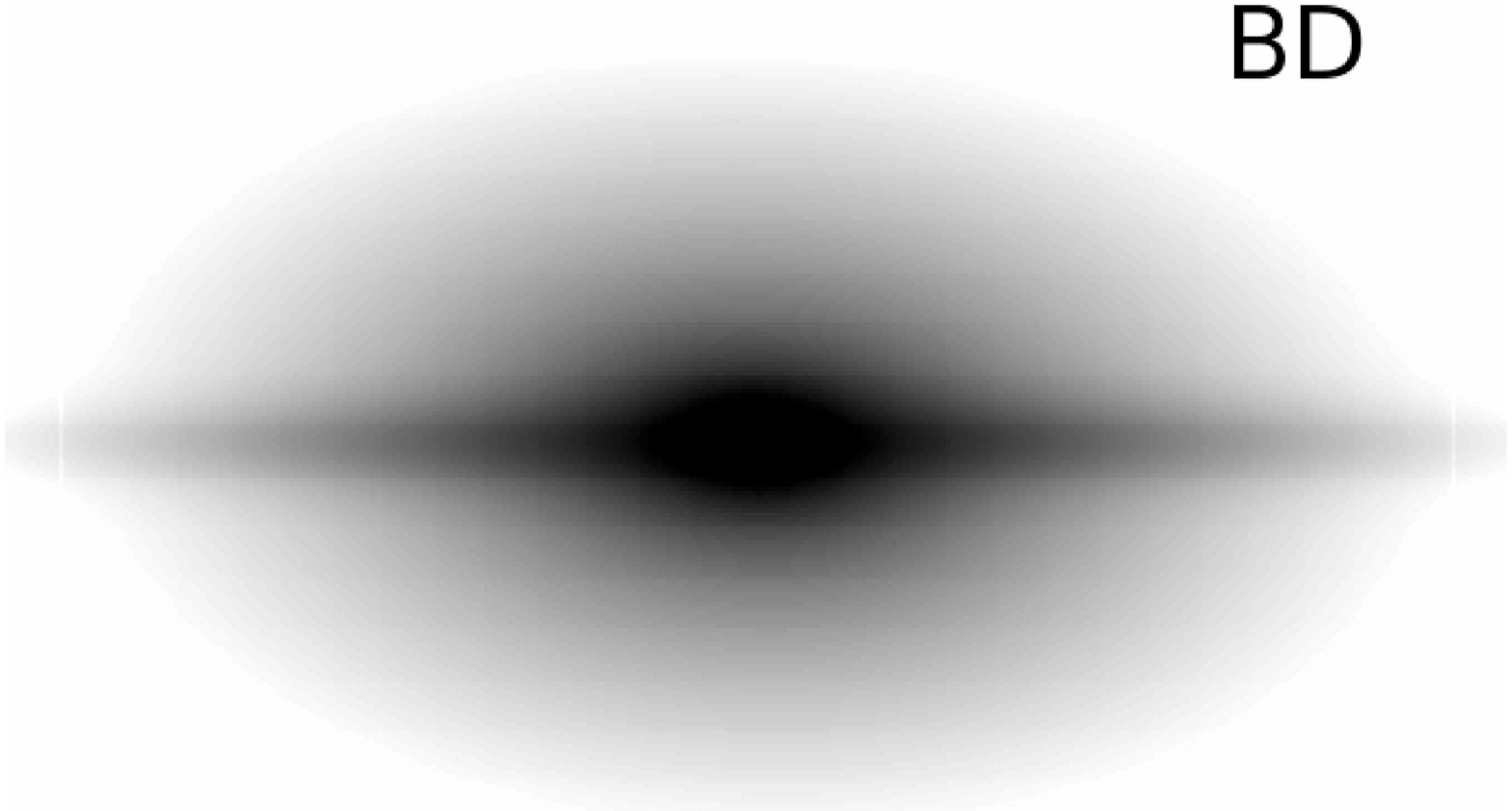}
   \includegraphics[keepaspectratio=true,width=5cm,clip=true]{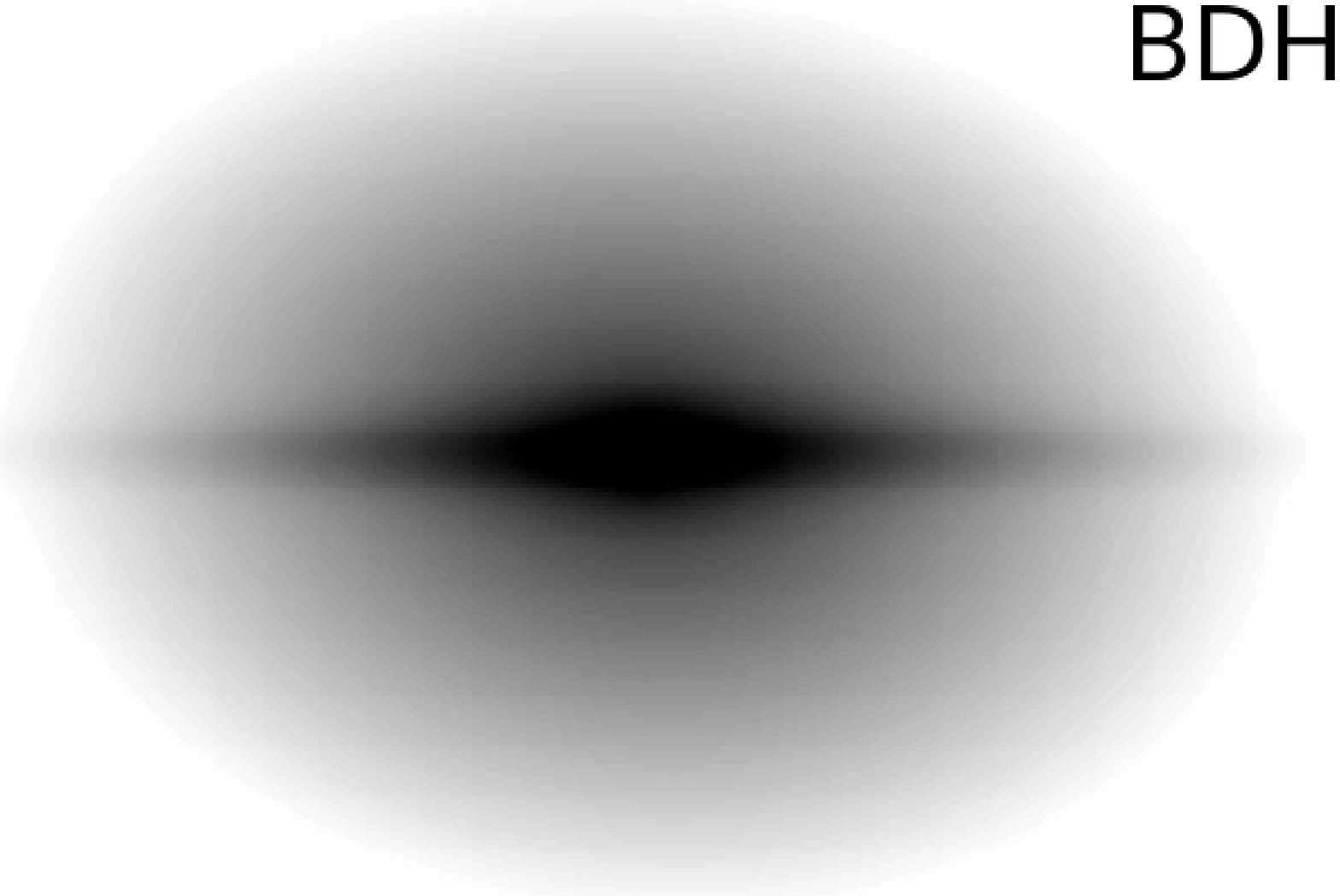}
   \includegraphics[keepaspectratio=true,width=5cm,clip=true]{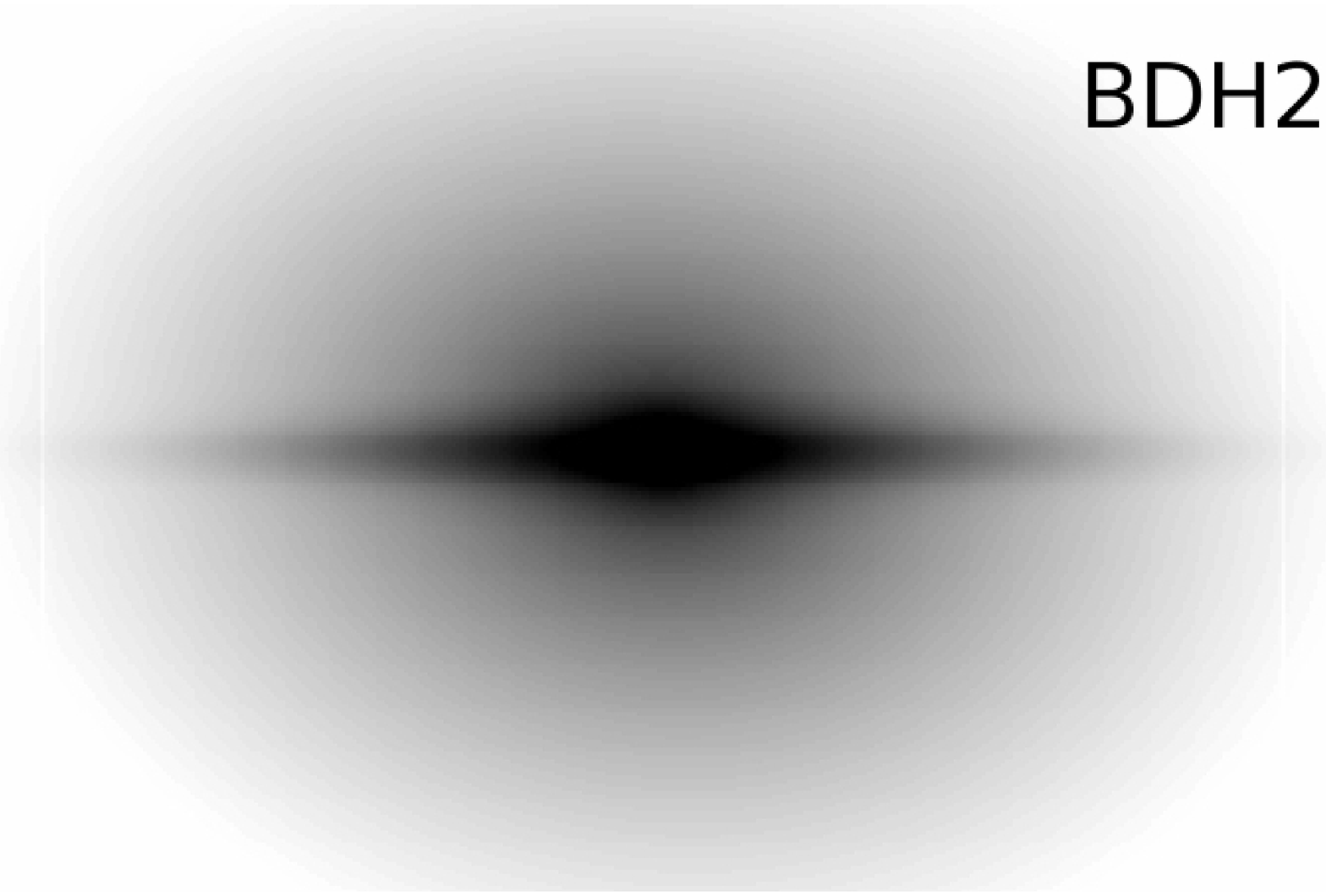}\\
   \includegraphics[keepaspectratio=true,width=5cm,clip=true]{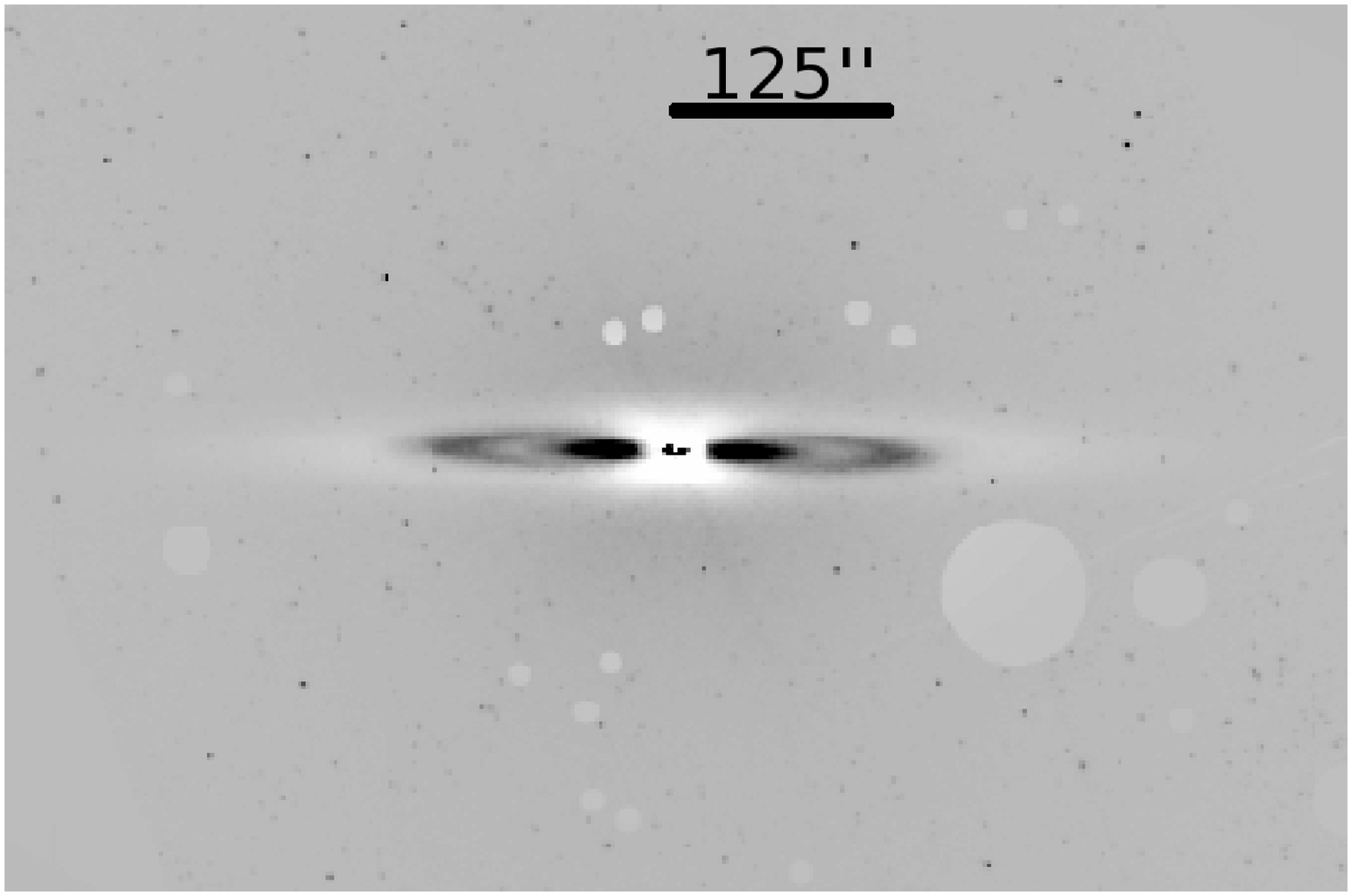}
   \includegraphics[keepaspectratio=true,width=5cm,clip=true]{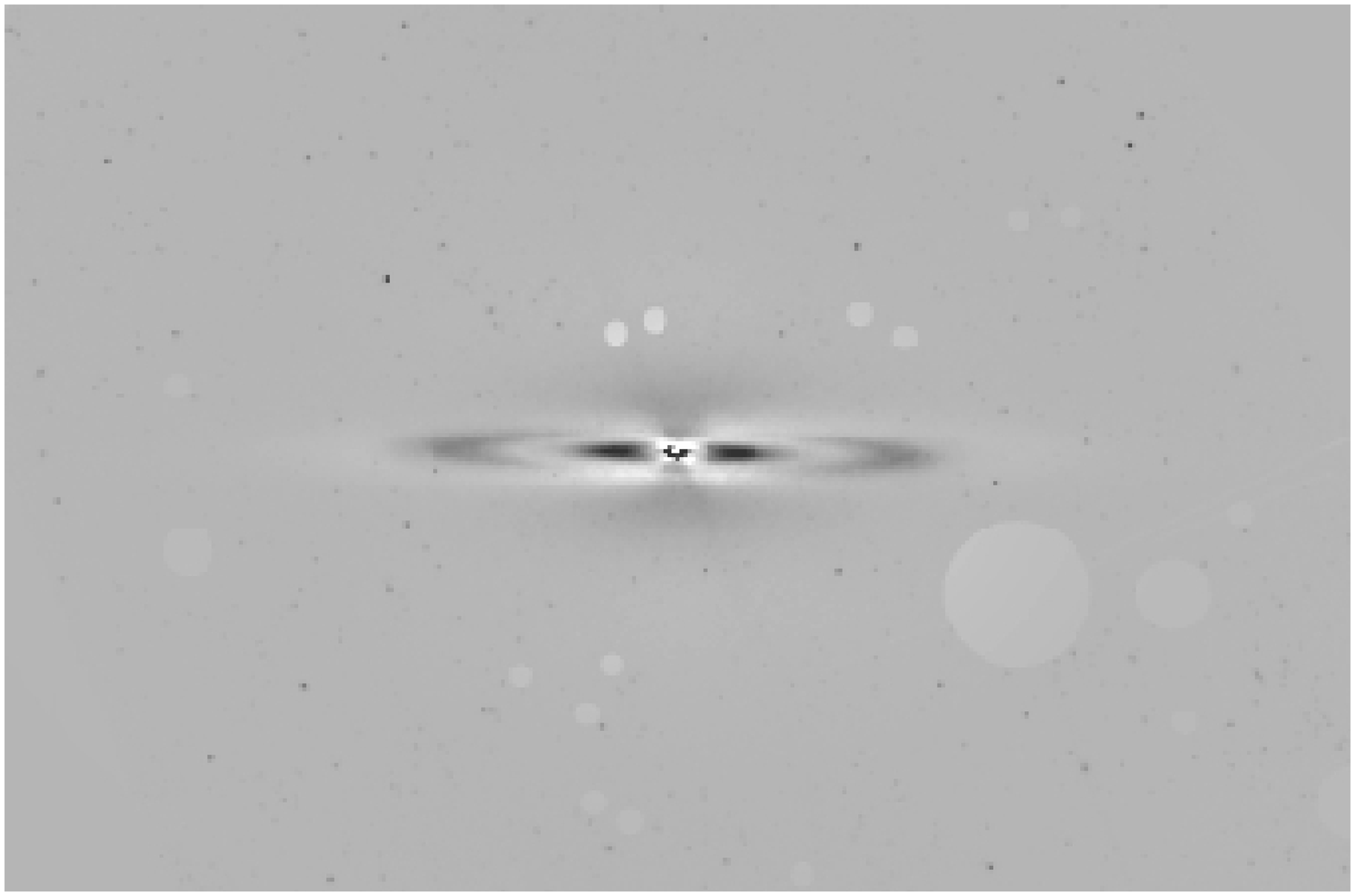}
   \includegraphics[keepaspectratio=true,width=5cm,clip=true]{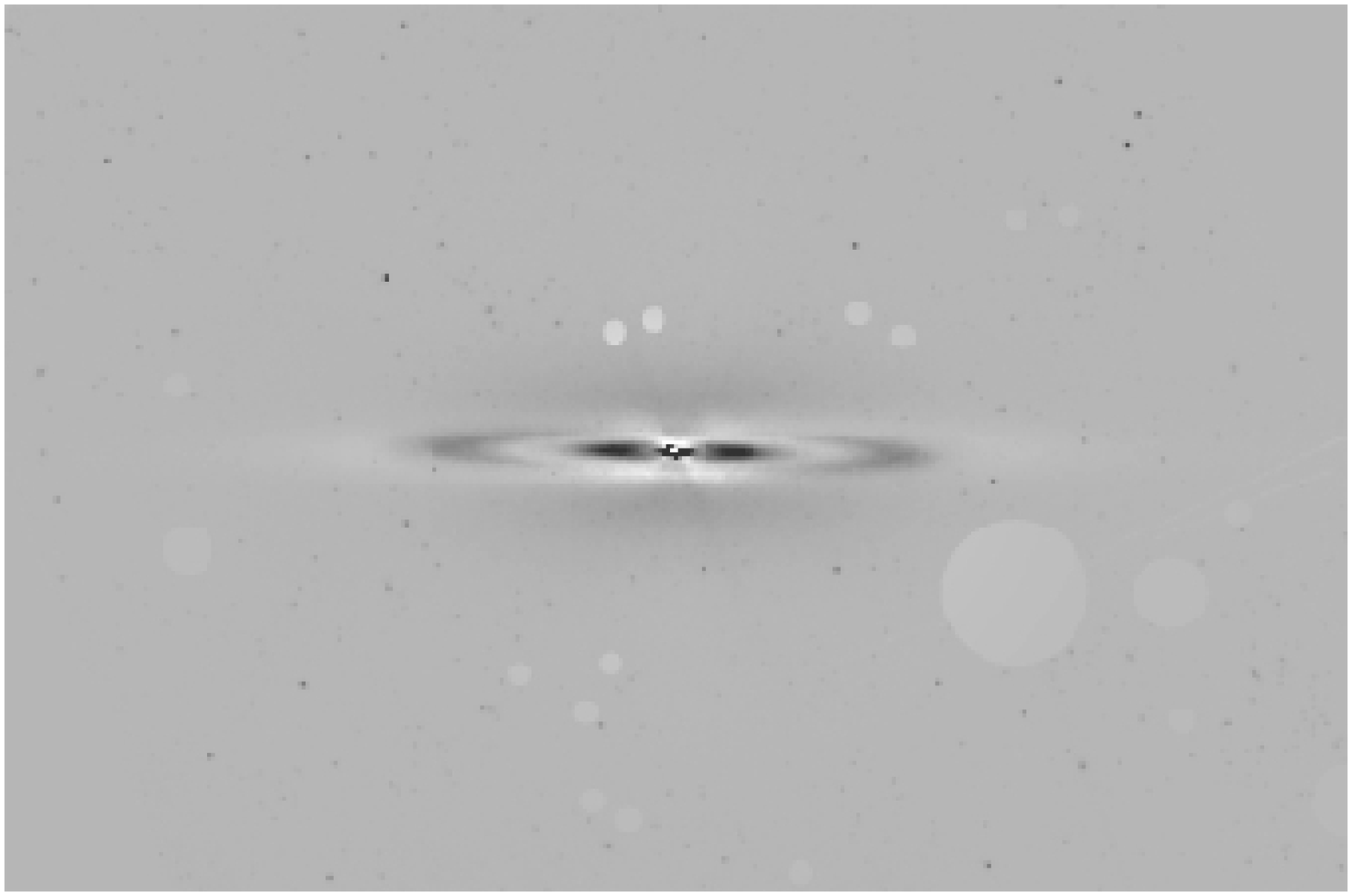}\\
   \includegraphics[keepaspectratio=true,width=5cm,clip=true]{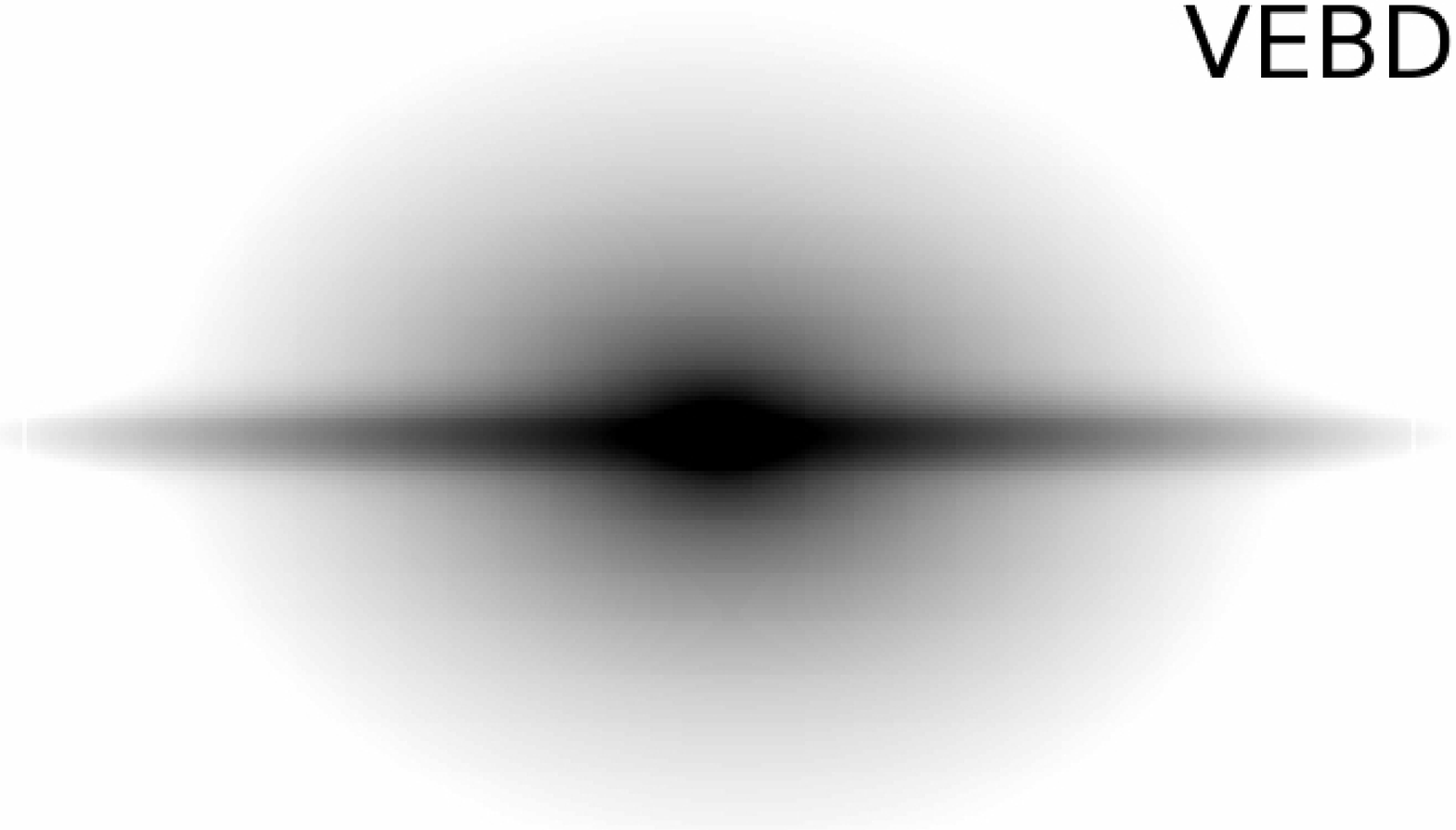}
   \includegraphics[keepaspectratio=true,width=5cm,clip=true]{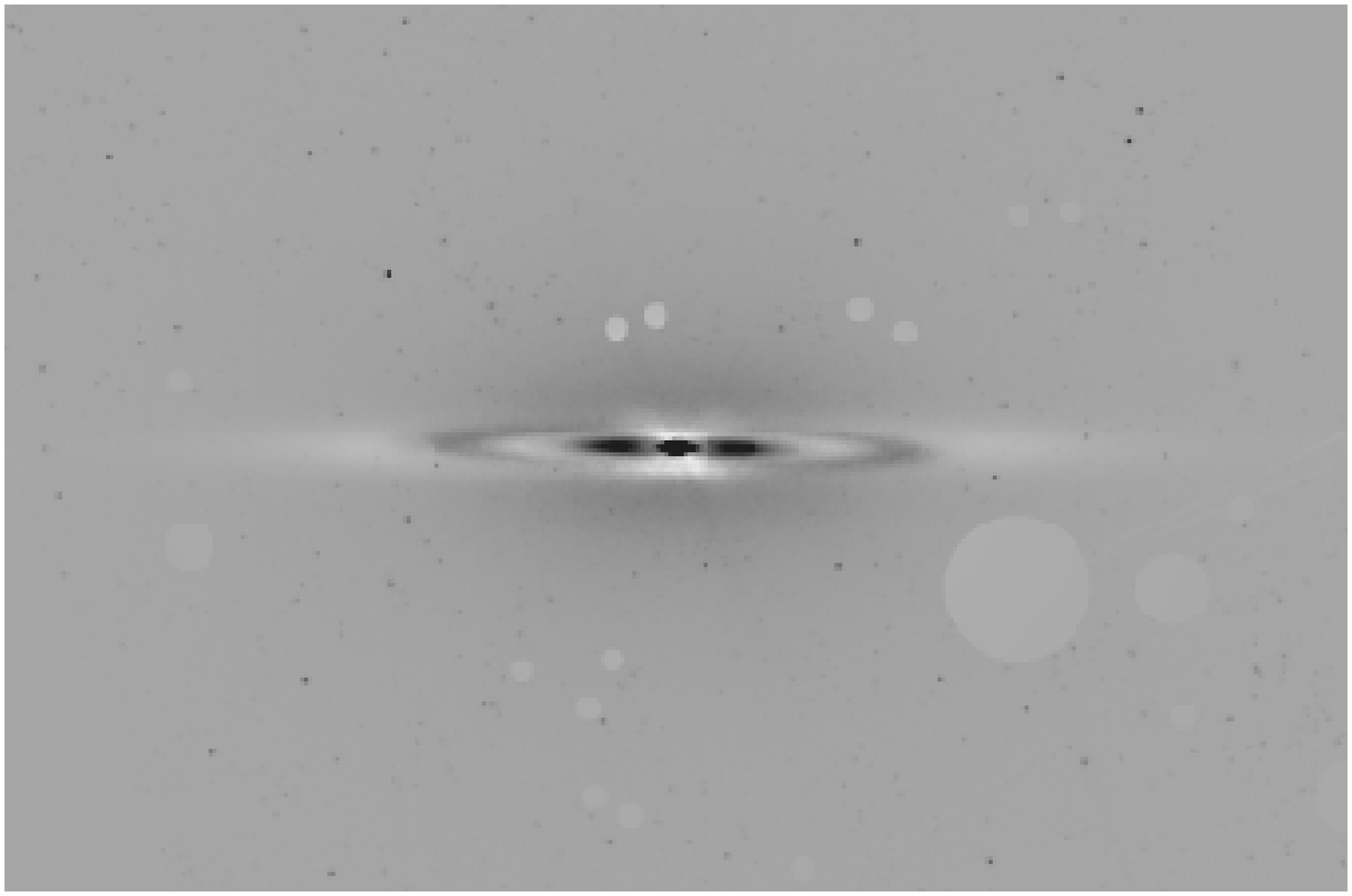}
   \caption{Images of the full {\sc budda} models (top) and the corresponding residual images (bottom) after subtraction of the models from the IRAC image. From left to right: bulge+disc model, bulge+disc and exponential halo model, bulge+disc and S\'ersic $n>2$ halo model. The top images are in the same scale and display levels as in Fig. \ref{fig:irac}. The residual images are all in the same scale and display levels. The one at left shows clearly that the bulge model is too luminous when the stellar halo component is not included. The two panels in the last row are the corresponding images for model {\sc vebd}, i.e. bulge with varying ellipticity+disc.}
   \label{fig:imgs}
\end{figure*}

For reasons that will be clear shortly, we also produced fits including an exponential halo (model {\sc bdh}) and a S\'ersic halo with S\'ersic index constrained to be larger than 2 (model {\sc bdh2}). The other free halo parameters are: effective radius, effective surface brightness and ellipticity (the position angle is the same as for the bulge). In addition, a fourth model was produced, with bulge and disc components only, but in which the bulge ellipticity varies with radius (model {\sc vebd}). The results from the fits are shown in Table \ref{tab:result} and Figs. \ref{fig:profs_bd} to \ref{fig:imgs}.

\begin{table*}
\centering
\begin{minipage}{115mm}
\caption{Results from the decompositions. Model {\sc bd} has bulge and disc only. Model {\sc bdh} includes an exponential halo, while model {\sc bdh2} includes a S\'ersic halo with S\'ersic index constrained to be larger than 2. Finally, model {\sc vebd} has a bulge with varying ellipticity and a disc. Models {\sc s1} to {\sc s3} are discussed in Sect. \ref{sec:minor} and are fits to the galaxy spheroid only. Model {\sc s1} is a single S\'ersic spheroid, model {\sc s2} is a single S\'ersic spheroid with ellipticity fixed at 0.1, and model {\sc s3} is a single S\'ersic spheroid with varying ellipticity as in model {\sc vebd}.}
\label{tab:result}
\begin{tabular}{@{}lccccccccccc@{}}
\hline \hline
Model & $h$ & $z$ & $n$ & $r_e$ & $\epsilon$ & $r_{e,{\rm h}}$ & $\epsilon_{\rm h}$ & B/T & D/T & H/T & $\chi^2$ \\
(1) & (2) & (3) & (4) & (5) & (6) & (7) & (8) & (9) & (10) & (11) & (12) \\
\hline
{\sc bd}     & 61 & 18 & 3.9$\pm$0.4 & 71 & 0.42 & --     & --     & 0.77 & 0.23 & -- & 5.9 \\
{\sc bdh}   & 41   & 17 & 1.9$\pm$0.2 & 10   & 0.46 & 130 & 0.28 & 0.13 & 0.35 & 0.52 & 2.9 \\
{\sc bdh2} & 44   & 15 & 2.0$\pm$0.1 & 9   & 0.41 & 168 & 0.23 & 0.09 & 0.30 & 0.61 & 3.5 \\
{\sc vebd} & 56   & 14 & 4.3$\pm$0.5 & 69   & -- & -- & -- & 0.68 & 0.32 & -- & 5.5 \\
\hline
{\sc s1} & --   & -- & 3.3$\pm$1.5 & 80   & 0.38& -- & -- & -- & -- & -- & -- \\
{\sc s2} & --   & -- & 2.5$\pm$0.7 & 79   & 0.10 & -- & -- & -- & -- & -- & -- \\
{\sc s3} & --   & -- & 2.5$\pm$0.8 & 81   & -- & -- & -- & -- & -- & -- & -- \\
\hline
\end{tabular}
Column (1) shows the model fitted. Columns (2) and (3) are the disc scale-lenght and scale-height, respectively. Columns (4) to (6) show the bulge S\'ersic index, effective radius and ellipticity, respectively. Columns (7) and (8) display the halo effective radius and elliptictity, respectively. Columns (9) to (11) are, respectively, bulge-to-total ratio, disc-to-total ratio and halo-to-total ratio. Finally, column (12) is the reduced $\chi^2$ of the fit. All scales are in arcseconds and the uncertainties in the bulge S\'ersic index are also shown. Check how the bulge ellipticity varies with radius in model {\sc vebd} at the middle panel of Fig. \ref{fig:geo}.
\end{minipage}
\end{table*}

The left panel in Fig. \ref{fig:profs_bd} shows surface brightness radial profiles of NGC 4594 and model {\sc bd}, obtained through ellipse fits to the isophotes of both images, using the {\sc ellipse} task in {\sc iraf}. If considered alone, it indicates that a good fit can be obtained with bulge and disc components only.

However, in the case of edge-on galaxies, using ellipse fits to study the light distribution in galaxies can be misleading. In most galactocentric distances, the isophotes in an edge-on galaxy can result from more than one component. In the case of Sombrero, it is easy to see that, except from outermost isophotes, many isophotes are drawn from light coming from the disc, bulge and halo (see Fig. \ref{fig:irac}). Thus, in this case, surface brightness radial profiles from ellipse fits are unsuitable to assess how light is distributed amongst different structural components, and hence to carefully judge the quality of a 2D fit, since at each given radius such profiles respond in a complex fashion to how the different components influence the isophotes. Note that in the case of face-on or less inclined galaxies, although at any given isophote there is likely a contribution from more than one structural component only, such components are usually strongly dominant at specific radii intervals, which overcomes this difficulty, and render surface brightness radial profiles from ellipse fits to isophotes suitable.

Thus, in order to more properly evaluate how light is distributed amongst the different structural components in the case of Sombrero, we made cuts along the major and minor axes of the disc to produce another set of surface brightness radial profiles (see right panel of Fig. \ref{fig:profs_bd} -- the minor axis cuts are discussed in Sect. \ref{sec:minor}). In contrast to profiles built via isophotes, in this second set of profiles, each radius correspond essentially to a unique galactocentric radius. Thus, in this case, each component dominates over the others at different radii intervals. Now one can see that model {\sc bd} is a relatively poor fit to the outer parts of NGC 4594. 

In the right panel of Fig. \ref{fig:profs_bd}, the galaxy profile exhibits two important breaks (indicated by circles): the first, at $r\approx28"$, marks the limit inward to which the bulge dominates over the disc, while the second at $r\approx215"$, marks the limit outward to which the outer spheroid dominates over the disc. Note that the alternative interpretation in which the second break indicates where the bulge starts to dominate again over the disc is weakened by the fact that model {\sc bd} does not fit the outer profile well, even with a bulge with S\'ersic index $n=3.9$. This is due to the fact that the outer profile can be very well described by an exponential function. Between these two breaks is where the disc dominates, and, in this region, one can see two bumps that are likely caused by two different stellar rings, which will be discussed further below. In addition, the middle panel in Fig. \ref{fig:geo} shows that the ellipticity of the outer galaxy isophotes is substantially lower than those of the bulge component in model {\sc bd}, which has an ellipticity $\epsilon=0.42$. Furthermore, the corresponding residual image in Fig. \ref{fig:imgs} shows clearly that the bulge component in model {\sc bd} is also too luminous in the inner parts. Note that fitting an image where the outer ring is masked out produces very similar results (see the red lines in the right panel of Fig. \ref{fig:profs_bd}). This test was done by masking only the outer ring component, and is different from the test performed in Sect. \ref{sec:minor}, in which the entire disc, ring and central sub-structures are masked out. In a 1D, profile fit, the ring could produce a systematic effect, but because Sombrero is nearly edge-on, and the fits shown here are 2D fits, such systematic effect is absent, as the ring occupies a very small fraction of the pixels in the whole image being used by the code to produce the fit. It is also worth noting that {\sc budda} produces fits to the average, underlying distribution of light from each component \citep[see e.g.][]{GadBaeFal10,ScaGadJon10}. Further, its $\chi^2$ minimization algorithm does search for the global $\chi^2$ minimum through simulated annealing, avoiding to be trapped in a local $\chi^2$ minimum.
In the Appendix, we show that including two ring components in the modeling does not lead to significantly different results.

We thus added an exponential halo\footnote{We use the word `halo' to describe an additional component, whose presence is justified from a structural viewpoint. We caution the reader that this does not necessarily imply the existence of a distinct constituent in terms of stellar population content (e.g. stellar ages and metallicities).} to produce model {\sc bdh} (Fig. \ref{fig:profs_bdh}). The ellipse fits analysis suggests in this case an equally good fit, as compared to model {\sc bd}. However, the cuts along the major axis, which, as argued above, is the proper tool to use in the case of Sombrero (together with cuts along the minor axis -- see below), indicate that model {\sc bdh} is an improved fit. The maximum in the residual profile, excluding the rings, drops by a factor of $\sim2$. The reduced $\chi^2$ also drops by a factor of two. Note that this improvement of a factor 2 happens when moving from 10 free parameters to 13 free parameters, i.e. an increase of 30\% in the number of free parameters.

For the sake of exploring different alternatives, in model {\sc bdh2} we tried a S\'ersic halo, forcing a S\'ersic index larger than 2, as otherwise it would drop to about 1, as in model {\sc bdh}. The best fit S\'ersic index for the halo in this case is 2.4. Figure \ref{fig:profs_bdh2} shows that the fit in this case is not good in the outer parts, as the halo is too luminous.

Figure \ref{fig:geo} shows clearly how adding a halo to the model results in a much better fit to the geometrical properties of NGC 4594, as compared to model {\sc bd}. The model images in Fig. \ref{fig:imgs} show that model {\sc bdh} reproduces better the top panel in Fig. \ref{fig:irac}. One reaches a similar conclusion analysing the residual images in Fig. \ref{fig:imgs}. The residual image from model {\sc bd} clearly shows that the bulge model is too luminous in the centre when the halo is not accounted for. As discussed above, this cannot be an effect arising from not including an AGN component, as this component, if at all necessary, is too faint. It is also unlikely to be a problem of PSF fitting, since we tried two different PSF models, which yielded similar results, and the area where the residual caused by the bulge is excessive has a size of almost $\approx$1', i.e several times larger than the PSF FWHM, which we measured as 1.9''. It is thus clear that model {\sc bdh} is a better fit, even if not fitting completely the galaxy image, for not having ring components. The residual images in Fig. \ref{fig:imgs} show clearly a ring in the disc of NGC 4594. Furthermore, it also shows an elongated central sub-structure, which could be an inner ring or disc. This is different from the relatively faint nuclear disc already found by \citet{Bur86}, with a size of $\approx$15'', which can also be seen in our residual images, but does not produce a clear imprint in the radial surface brightness profile of the galaxy. Both such structures induce the bumps in the surface brightness radial profile of NGC 4594, seen for instance in the right panel of Fig. \ref{fig:profs_bdh}, in the region where the disc dominates the profile. It should be noted that these residual images have a very narrow display stretch, in order to emphasize discrepancies between the model fitted and the galaxy, as well as existing sub-structures. Further, it is worth noting that, because the fits shown here are 2D fits, where the whole image is taken into account, as opposed to 1D profile fits, the residual images are also an appropriate tool to verify the goodness of the fits.

The variation of ellipticity with radius in Sombrero, seen in the middle panel of Fig. \ref{fig:geo}, suggests another, simpler possibility. A model with bulge and disc only, but where the bulge ellipticity varies with radius, could perhaps produce a fit to the galaxy at least as good as model {\sc bdh}. \citet{Ryd91} studied the geometry of ellipsoids with ellipticity varying with radius in different projections. Her figures 3 and 4 show a radial behaviour of ellipticity that resembles what we observe for Sombrero.  We thus tailored {\sc budda} to allow for a bulge with varying ellipticity $\epsilon$, in order to produce model {\sc vebd}. In this model, thus, as in model {\sc bd}, the word `bulge' refers to the whole spheroidal structure seen in the galaxy, from the central bulge to the outer halo. However, to not introduce an excess of degenerated solutions, the way $\epsilon$ varies with radius is fixed. The ellipticity variation shown in Fig. \ref{fig:geo} gives us information on how this should be fixed, but one has to keep in mind that the variation measured through ellipse fits corresponds to the whole galaxy, not the bulge component alone. In order to evaluate the intrinsic ellipticity variation of the bulge component alone, one can look at the outermost radii, where the contribution of the disc component drops substantially. The ellipticity profile in Fig. \ref{fig:geo} shows that outside a radius of about 250 arcsec, where the outer spheroid dominates over the disc, $\epsilon$ drops more or less linearly from $\approx0.33$ to $\approx0.1$. Inside this radius, the ellipticity of the isophotes reflects that of both disc and bulge. There is in fact almost a plateau in $\epsilon$ around this radius of 250 arcsec. A natural choice is thus to keep $\epsilon=0.33$ for the bulge component inside this radius. This choice is also supported by the fact the ellipticity of the innermost isophotes in the galaxy is 0.41, and therefore relatively close to 0.33. The green line in the middle panel of Fig. \ref{fig:geo} shows how $\epsilon$ varies with radius for the bulge component only in model {\sc vebd}. Note that this choice also leads to a better fit using model {\sc vebd}. If one leaves the bulge ellipticity rising linearly until the centre, the resulting model is too eccentric. What happens is that the disc, because it is almost edge-on, has an ellipticity of about 0.9\footnote{Note that our disc model is viewed perfectly edge-on, and in this case the concept of ellipticity is not applicable as it is for face-on or inclined discs. The disc ellipticity is simply not a parameter in our models. The value of 0.9 is a reference value, and it comes from the measure of disc inclination by \citet{emsbacmon96}. The disc scale length and scale height determine how thin the disc is, and one sees that their ratio varies amongst our models.}. The spheroid being rounder helps dilute this eccentricity, less so where the disc dominates. If the spheroid is also eccentric this dilution is reduced. In fact, one can see that even with a bulge model with a maximum ellipticity of 0.33, the dilution is too weak to reproduced the galaxy ellipticity profile (blue solid line in the middle panel of Fig. \ref{fig:geo}).

Figure \ref{fig:profs_veb} shows the results of the fit from model {\sc vebd}. It shows that, although the fit looks better than the one from model {\sc bd}, model {\sc bdh}, with its $n\approx2$ bulge and an exponential halo, still provides a better fit than model {\sc vebd}, with a bulge with varying $\epsilon$ and $n\approx4$, and no halo. In fact, the reduced $\chi^2$ from model {\sc vebd} is similar to that of model {\sc bd}, i.e. a factor about 2 worse than model {\sc bdh}. Further, although the ellipticity of the full model {\sc vebd} reproduces the more or less linear fall with radius, it does so with an offset from the values of ellipticity measured in the galaxy (Fig. \ref{fig:geo}). It might be surprising that model {\sc vebd} does not do much better, given its degrees of freedom. The crucial factor here appears to be the shape of the profile. Given that the outer part of the galaxy profile is quite strictly exponential, the fitting of a single spheroid leading to a S\'ersic index of about 4 results in a bad fit. Further below, we will discuss the possibility that the $n\approx2$ bulge in model {\sc bdh} is actually a bar seen end-on. This bar might be the one component which is leading model {\sc bd} to a relatively bad fit. Figure 1 in \citet{BurAth05} shows unambiguously that an end-on bar remarkably resembles a classical bulge from a purely morphological viewpoint.

\begin{figure}
   \centering
   \includegraphics[keepaspectratio=true,width=8.2cm,clip=true]{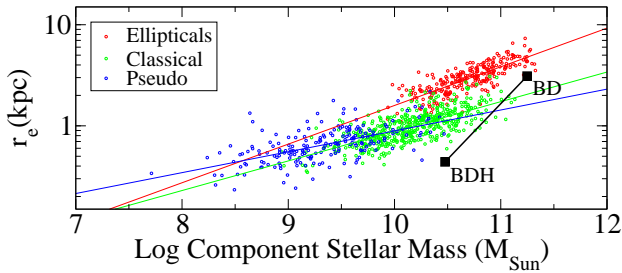}
   \caption{Mass-size relations for elliptical galaxies, classical bulges and pseudo-bulges from data in \citet{Gad09b}. The position of the Sombrero bulge in both models {\sc bd} and {\sc bdh} is marked. The spheroid in Sombrero as given by model {\sc bd} is larger and more massive than any of the bulges. In fact, it is as massive as the most massive elliptical galaxies, and falls near the mass-size relation of ellipticals, rather than that of (classical) bulges. On the other hand, the bulge obtained in model {\sc bdh}, although somewhat small for its mass, falls near the relation drawn for classical bulges, although the locus it occupies in this relation is also not unusual for a disc-like, pseudo-bulge.}
   \label{fig:scale}
\end{figure}

The most striking result from this analysis is thus that the bulge S\'ersic index of the Sombrero galaxy drops from 3.9 to 1.9, when one does a fitting of the IRAC image including the stellar halo in the model. Concomitantly, $B/T$ drops from 0.77 to 0.13\footnote{\citet{GraWor08} discussed how early studies forcing the bulge S\'ersic index at 4 resulted in overestimated bulge-to-total ratios.}, while the bulge effective radius $r_e$ drops from 3.1kpc to 0.44kpc. These drops in $B/T$, $n$ and $r_e$ make the bulge in Sombrero look much more typical, since the average values for Sa galaxies are $B/T=0.31$ (with a 1$\sigma$ scatter lower limit of 0.15), $n=2.6$ and $r_e=0.7$kpc \citep[][from $K$-band data -- see also \citealt{Gad09b}]{GraWor08}. In this context, it is worthy to point out that \citet{temten06} found a value of $B/T=0.15$ through sophisticated dynamical modeling of the mass distribution in NGC 4594, including a halo component, in very good agreement with our model {\sc bdh}. Note that the value obtained for the bulge effective radius when the halo is not included in the fit is almost a factor 5 larger than the typical value for other Sa galaxies. Another way of looking at this is put in Fig. \ref{fig:scale}. It shows that the spheroid in NGC 4594, if treated as a single component, resembles, from a structural point of view, more an elliptical galaxy than the bulge of a disc galaxy. In this figure, we show the mass-size relations obtained in \citet{Gad09b} for nearly 1000 elliptical galaxies, classical bulges and pseudo-bulges, drawn from a complete parent SDSS sample, with a typical redshift of 0.05. The spheroid in Sombrero as given by model {\sc bd} is larger and more massive than any of the bulges found in that sample. In fact, it is as massive as the most massive elliptical galaxies, and falls near the mass-size relation of ellipticals, rather than that of (classical) bulges. In this context, the scenario entertained by \citet{emsbacmon96}, in which the spheroid forms after the dissolution of a bar, is highly improbable, given that the typical bar is about two orders of magnitude less massive than the spheroid in Sombrero \citep[see e.g.][]{Gad11}. On the other hand, the bulge obtained in model {\sc bdh}, although somewhat small for its mass, falls near the relation drawn for classical bulges, although the locus it occupies in this relation is also not unusual for a disc-like, pseudo-bulge.

In summary, our structural analysis shows that the most appropriate model for Sombrero is model {\sc bdh}, with a spheroid comprising an inner bulge and an outer halo. Nevertheless, we cannot rule out the possibility that the spheroid is in fact an elliptical galaxy (as suggested in Fig. \ref{fig:scale}), that, unlike most elliptical galaxies, displays intricate structural properties, due to the presence of a massive embedded disc, and the possibility that this disc have formed a bar and thus experienced effects from secular evolution. This opens up the possibility that the current view on the Sombrero spheroid as a single entity, a large classical merger-built bulge, is mistaken. These possibilities will be discussed in more detail in the next section.

\subsection{Fits to spheroid only}
\label{sec:minor}

\begin{figure}
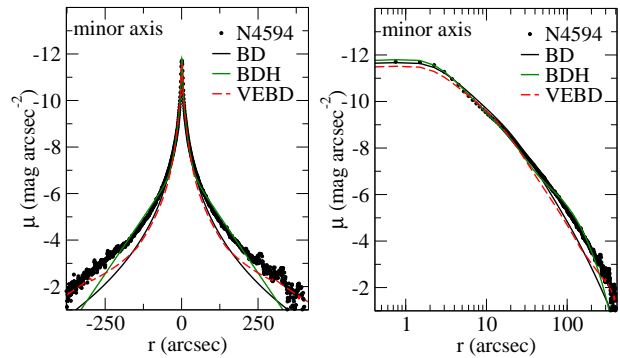

   \centering
   \includegraphics[keepaspectratio=true,width=4cm,clip=true]{N4594vec_minor_10.eps}
   \includegraphics[keepaspectratio=true,width=4cm,clip=true]{N4594vec_minor_10_2.eps}
   \caption{Left: minor axis surface brightness profiles of NGC 4594 and 3 of the models fitted, as described. Right: Same profiles with with the radial axis in logarithmic scale, to emphasize the inner parts.}
   \label{fig:minor}
\end{figure}

\begin{figure}
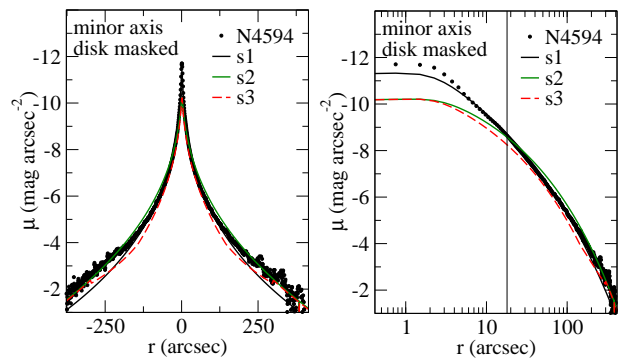

   \centering
   \includegraphics[keepaspectratio=true,width=4cm,clip=true]{N4594vec_minor_10_masked.eps}
   \includegraphics[keepaspectratio=true,width=4cm,clip=true]{N4594vec_minor_10_masked_2.eps}
   \caption{Same as Fig. \ref{fig:minor}, but for the models with only a spheroid. Model {\sc s1} is a single S\'ersic spheroid. Model {\sc s2} is a single S\'ersic spheroid with ellipticity fixed at 0.1. Model {\sc s3} is a single S\'ersic spheroid with varying ellipticity as in model {\sc vebd}. The vertical line marks one disc scale-height, within which the galaxy image was masked out.}
   \label{fig:minor_s}
\end{figure}

Another important way to address the nature of the spheroid in Sombrero is by looking at minor axis profiles, where the influence of the inner sub-structures, disc and rings is minimized. In Fig. \ref{fig:minor} we plot the minor axis profiles of the galaxy and models {\sc bd}, {\sc bdh} and {\sc vebd}. One sees that model {\sc bdh} provides a better fit, although it does not fit well the outer parts. This is very likely because the halo has a fixed ellipticity of 0.28, somewhat larger than the ellipticity of the outermost parts of the galaxy, which is around 0.1. Model {\sc vebd} provides a better fit than model {\sc bd}, attesting the importance of having an ellipticity varying with radius. At the inner parts, models {\sc bd} and {\sc bdh} provide equally good fits, whereas model {\sc vebd} is slightly off.

In this context, we produced further fits by excluding all sub-structures apart from the spheroid. This was done by masking a rectangular area centred at the galaxy center, and with dimensions 36'' $\times$ 430'', along the disc major axis. The minor axis dimension is twice the disc scale height (as in model {\sc bd}), and the major axis dimension was chosen to allow only for the part dominated by the spheroid, as seen in the galaxy profile (see e.g. Fig. \ref{fig:profs_bd}).

The new fits include a S\'ersic spheroid only, with the S\'ersic index left free. Model {\sc s1} has free ellipticity, model {\sc s2} has a fixed ellipticity of 0.1, i.e. reproducing the outermost parts of the galaxy ellipticity radial profile (see Fig. \ref{fig:geo}), and finally, model {\sc s3} has varying ellipticity, as in model {\sc vebd}. The resulting parameters are shown in Table \ref{tab:result}. Minor axis profiles from these fits are shown in Fig. \ref{fig:minor_s}. These new models generally provide better fits to the outer parts of the galaxy than the models in the previous sub-section, but fail in the inner parts. This evidently reflects the design of this exercise and the influence the disc component has in the global fits of the previous sub-section.

Model {\sc s1} is the one that fails the most in the outermost parts. Model {\sc s2} is overall quite good but fails at intermediate radii, i.e. from about 50'' to 100''. Finally, model {\sc s3} can account for the change in ellipticity in the outermost parts, but generally does not result in a very good fit. This exercise illustrates the fact that even excluding from the fit all sub-structures apart from the galaxy spheroid, and even allowing for varying ellipticity, Sombrero's spheroid is not well described by a single S\'ersic component. This justifies adding an extra component apart from the main bulge and disc components. Furthermore, the left panels in Figs. \ref{fig:minor} and \ref{fig:minor_s} clearly show that the outermost surface brightness profile of Sombrero follows very well an exponential profile, as the halo in model {\sc bdh}, and thus cannot be well reproduced by a S\'ersic function with high S\'ersic index. In fact, as shown in Table \ref{tab:result}, model {\sc s1} has $n=3.3$, while models {\sc s2} and {\sc s3} have $n=2.5$. This suggests that the change of ellipticity is related to a flatter surface brightness profile, since models {\sc s2} and {\sc s3} not only have lower S\'ersic indices, but also reproduce better the outer lower ellipticities seen in the galaxy. This is a further indication of the presence of an extra spheroidal component, with flatter mass distribution and lower ellipticity.

\section{The intriguing nature of Sombrero's spheroid}
\label{sec:disconc}

\subsection{Evidence from structural analysis}

The work done in the previous section is a compelling illustration of the fact that structural analysis of galaxies through image decomposition is a powerful tool, yet rather a complex affair. It demonstrates that results can significantly depend on details that show up only through better data and/or careful scrutiny. In fact, 1D profile fits alone can also be misleading, as shown in Fig. 7 of \citet{Gra01}. Our structural analysis indicates that the spheroid in Sombrero comprises two main components: an inner bulge and an outer halo.

\citet{benbucdal06} presented a similar study as ours using the same IRAC image. They did not include a halo component in their fits. However, their focus was on the dust distribution in Sombrero, and so they concentrated their efforts on the 24$\mu$m image, where, as they stated, the ``bulge almost disappears completely''. Although they showed a residual image from the fit to the 24$\mu$m image, they did not show intensity radial profiles, making it somewhat more difficult to assess how good was their fit. In particular, they did not show a residual image from their model to the 3.6 $\mu$m image, which basically traces only stars, and where the spheroid is very conspicuous. We also note that they fixed the S\'ersic index of their bulge model to $n=4$. Their $B/T$ for the 3.6$\mu$m image is 0.77, i.e. the same as we find in model {\sc bd}. \citet{jargebshe11} also presented a similar study, putting together HST and ground-based images. They used a non-parametric form to fit the bulge profile, plus 3 discs, and found $B/T=0.73$. They stated that their bulge profile ``could in principle be a combination of a S\' ersic bulge plus exponential halo''.

If the low value of the bulge S\'ersic index in the Sombrero galaxy is correct -- as the results in this study suggest since model {\sc bdh} gives us the best fit -- and if Sombrero has indeed a classical bulge, it is an example of how using the bulge S\'ersic index as a discriminator between classical and disc-like bulges can lead to wrong bulge classifications. This is 
only surprising because NGC 4594 is a prototype for classical bulges. Alternatively, the lower value for $n$ found here could mean that the Sombrero galaxy has actually a disc-like bulge, rather than a massive classical bulge, which would be truly surprising.

Two of the previous arguments in favour of the Sombrero bulge being a classical bulge are now weakened. Its S\'ersic index is not as high as thought, and it does not extend out from the plane of the disc as much as when the halo was considered part of the bulge (although even the bulge in model {\sc bdh} is not as flat as the disc). On the other hand, the lack of substantial star formation, as shown e.g. in GALEX observations \citep[see][]{gilboimad07}, suggests that Sombrero has a classical bulge.

Perhaps the most important criterion to identify a disc-like bulge comes from kinematics: in contrast to classical bulges, disc-like bulges have significant rotational support. \citet{kor88}, \citet{wagbendet89} and \citet{emsbacmon96} find strong rotational support in the inner regions of Sombrero. This would support the suggestion that the bulge in Sombrero is actually a disc-like bulge. In this case, the concept of the Sombrero bulge as a prototypical classical bulge is a result of the presence of its extraordinary stellar halo, whose full extent can only now be properly appreciated. Without knowing the true geometrical properties and light profile of this halo, it was immediately conceived as a massive, classical bulge. If we follow this reasoning, we conclude that, basically, what has been called `bulge' is actually a massive stellar halo. This halo, in contrast to the bulge, does not contribute considerably to the light/mass distribution at the galaxy central region.

The residual images in Fig. \ref{fig:imgs} clearly show a stellar ring, apart from another central stellar substructure, which could be a nuclear ring or disc. Suggesting evidence of these components have been published before \citep[see][]{waihylfre90,emsbacmon96}, but only with the IRAC image they stand out so clearly. An outer dust ring shows up beautifully in another Spitzer image \citep[][see also \citealt{vlabaeben08}]{kenarmben03}. Rings are usually, but not always, associated to bars \citep[see e.g.][]{grobutsal10}. An end-on bar in Sombrero can explain the central kinematics \citep{emsbacmon96}, and would be a natural explanation for the presence of the ring(s). \citet{emsfer00} went further and suggested that a {\em nuclear} bar could also be present. One cannot currently rule out the possibility that the bulge component fitted above in model {\sc bdh} is (at least partially) actually a bar seen end-on. This would be consistent with a photometric bulge with low S\'ersic index and yet no young stellar population, as bars are usually populated with old stars \citep[see e.g.][]{GaddeS06}. An end-on bar in Sombrero would also explain why the central isophotes are not as flat as the disc, as one would expect for disc-like bulges.

Using the value of $B/T=0.13$ found in model {\sc bdh}, the total galaxy stellar mass and the black hole mass vs. bulge mass relation given in \citet{HarRix04}, one finds that the expected black hole mass is $4.1\times10^7$M$_\odot$. This is in strong contrast to dynamical measurements given by \citet{korbenajh96}, which point out a black hole with mass about $10^9$M$_\odot$. More recent results also point to large black hole masses. \citet{jargebshe11} found a black hole mass of $6.6\times10^8$M$_\odot$. If one adds the halo luminosity to the bulge, the expected black hole mass rises up to $2.5\times10^8$M$_\odot$, which alleviates the discrepancy. If one uses the bulge mass in model {\sc bd}, the black hole mass derived is $3\times10^8$M$_\odot$. In this context, it is interesting to point out that the dynamical black hole mass measurement is also at odds with the black hole mass vs. velocity dispersion relation \citep[see][but see also \citealt{jargebshe11}]{beisarcor09}. Using the results in \citet{GulRicGeb09}, the expected black hole mass from the latter relation is $2.9\times10^8$M$_\odot$, although \citet{GraOnkAth11} argue for a value of $5\times10^8$M$_\odot$. It thus seems that either more sophisticated dynamical modeling is necessary to properly understand the central mass concentration in Sombrero, or it truly has an exceptionally massive black hole. Alternatively, the halo mass has also to be accounted for.

These results urge one to question thus how the halo in Sombrero did form, and why such structure is seen in some galaxies but not in all. For instance, NGC 4565 is also an edge-on galaxy, which was observed at the same wavelength also using Spitzer \citep{shereghin10}, more deeply than the Sombrero image used here, and yet it does not show such an extended halo. Conversely, NGC 5866, also observed by \citet{kenarmben03} does exhibit an extended halo. Early merger events are a likely formation mechanism for the Sombrero halo \citep{emsbacmon96}. Therefore, the properties of such haloes and their host galaxies, and how often they are present in galaxies, can provide constraints to physically characterise merger events, and elucidate how often and in which circumstances they occur. This is evidently an important issue for theories of structure growth in the universe.
In this context, it is worthy to point out that \citet{SeiGraJer07} and \citet{PieZibBra08} found that the haloes in BCGs (Brightest Cluster Galaxies) are better described with exponential luminosity profiles than with S\'ersic profiles with $n\sim4$, as in the case of Sombrero's halo. This also lends support to the arguments above that the main spheroidal in Sombrero is indeed a massive stellar halo.

Finally, one is left wondering on the impact of such extended haloes in our view of face-on galaxies. Since they might not be readily discernible in such cases, results from image decomposition can be substantially altered if they are not included in the model fitted, as seen above. Anti-truncated outer profiles can some times be a signature of the presence of a massive stellar halo, as pointed out by \citet[][see also \citealt{erwpohbec08}]{ErwBecPoh05}. Note that such a feature is clearly seen in the outer parts of Sombrero's radial surface brightness profiles shown in Sect. \ref{sec:struc}. The rightmost circle in the right panel of Fig. \ref{fig:profs_bd} marks the position where the anti-truncation begins.

\subsection{Evidence from further scaling relations}

As mentioned in the Introduction, NGC\,4594 is often regarded as a prototypical early-type disc galaxy. With its massive spheroid and the almost edge-on disc, it is usually classified as an Sa spiral \citep{deVdeVCor91}, although some authors prefer to consider it an S0 (e.g., \citealt{RhoZep04}, RZ04 hereafter).
The results presented in the previous sections, however, raise the intriguing possibility that the spheroid may not be a classical bulge, but to a large extent a stellar halo or an elliptical galaxy (see Fig.\,8). The latter possibility could in principle be considered just a matter of semantics if one takes the view that bulges are simply ellipticals surrounded by a prominent disc \citep[e.g.,][]{Ren99,KorKen04}, but Fig. \ref{fig:scale} shows that, at least at the high mass end, bulges and ellipticals follow different scaling relations. At fixed stellar mass, ellipticals tend to have larger effective radii than bulges, and it is noteworthy that Sombrero's spheroid as a whole lies much closer to the sequence defined by the former. While this does not necessarily preclude that they might share similar formation mechanisms, it clearly indicates that at least their formation histories must differ.

The resemblance of Sombrero's spheroid with elliptical galaxies has of course been noted earlier. \citet{HesPel93} pointed out that the mild colour and (absorption) line strength gradients, and the central kinematics, indicate such similarity. Indeed, the red colours of the spheroid inner regions gradually turn bluer with radius, matching those of the metal-rich globular cluster system (GCS) at about 5\,kpc from the centre \citep{SpiLarStr06}. This behaviour appears to hold out to large galactocentric radii, as \citet{MouSpi10} found that the metallicity distribution function (MDF) of stars in a field at about 18\,kpc from the centre peaks at the same value ([Fe/H] $\approx -0.5$) as the metal-rich GCS -- potentially indicating a link between the formation of both components. This peak is significantly more metal-rich than the halo of the Milky Way analog NGC\,891 ([Fe/H] $\approx -1$; \citealt{RejMouIba09}), but comparable to what is found in other elliptical galaxies (e.g. NGC\,5128; \citealt{HarHar02}). 

\begin{figure}
   \centering
   \includegraphics[width=0.45\textwidth,clip=true]{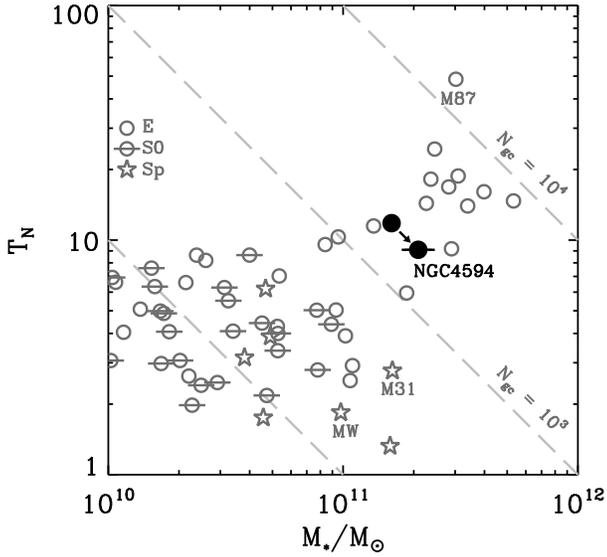}
      \caption{Globular cluster mass specific frequency of elliptical, lenticular and spiral galaxies in different environments (see text for details). NGC\,4594 has the largest GCS of all known disc galaxies, and occupies a locus that is more characteristic of intermediate-mass ellipticals. The black solid circle indicates the corresponding $T_{N}$ value if the whole GCS were associated just to the spheroid -- i.e., not taking the disc mass into account.}
         \label{sombrero_gcs}
   \end{figure}

The GCS in Sombrero perhaps provides the most compelling evidence of its spheroid's peculiar nature. It has long been known that NGC\,4594 possesses  the largest GC population ever found in a disc galaxy. RZ04 uncovered a GCS exceeding $N_{gc} = 1900$ globular clusters that extends out to $\sim$\,50\,kpc, with $\sim$\,40\% of them being metal-rich. Figure \ref{sombrero_gcs} shows the GC mass specific frequency\,\footnote{$T_{N} = N_{gc}/(M_{\star}/10^{9}\,{\rm M}_{\odot}$); \citet{ZepAsh93}.} for a compilation of M$_{\star} > 10^{10}$ M$_{\odot}$  elliptical, lenticular and spiral galaxies in different environments -- from field/groups \citep{SpiForStr08} to the Virgo cluster \citep{PenJorCot08}. With a factor $\sim$\,4 more GCs than M31, and inhabiting a low-density environment, it is clear that both the mass and $T_{N}$ of Sombrero are more characteristic of elliptical galaxies than of lenticulars or spirals. This plot also indicates that a major merger event between two M31-like spirals cannot reproduce the GC abundance of NGC 4594 -- not even that of the old, metal-poor population, which differs by a factor of about 3.5 between M31 and Sombrero. \citet{SpiLarStr06} pointed out that Sombrero has the largest \emph{bulge} specific frequency of red GCs among all spirals and thus its value is more characteristic of more massive galaxies. Moreover, they discovered that metal-poor GCs follow a colour-magnitude relation, making NGC\,4594 the first disc galaxy where such trend was observed. More recently, \citet{HarSpiFor10} showed that this mass-metallicity relation scales as $Z \propto L^{0.3}$, pretty much equivalent to what has been found in most massive ellipticals.

In summary, the spheroid in Sombrero has properties that differ significantly from those of classical bulges, including its mass-size relation; the richness of its GCS (and especially the metal-rich population); and the high metallicity of the outer halo when compared to other disc galaxies. 

\begin{figure}
   \centering
   \includegraphics[width=0.45\textwidth,clip=true]{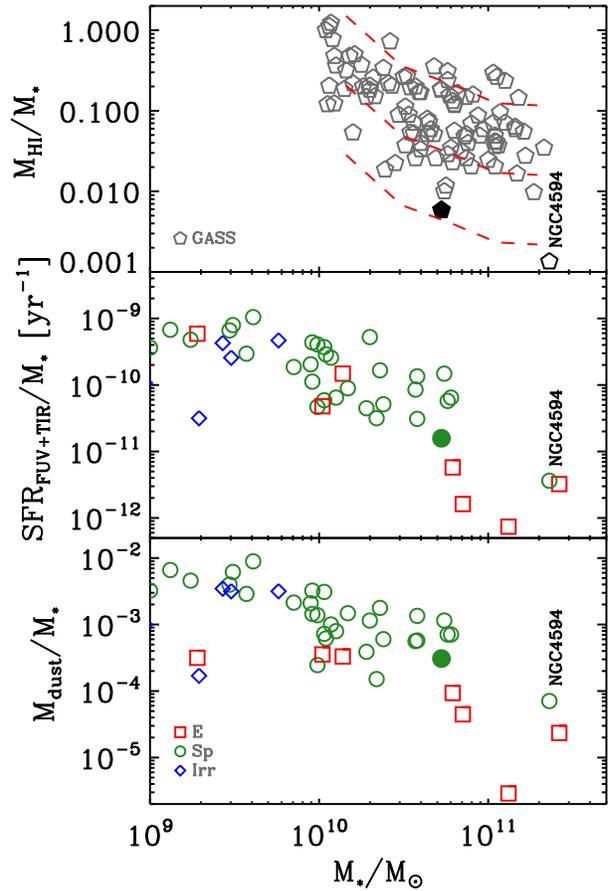}
      \caption{Scaling relations of NGC\,4594 as compared with those of other M$_{\star} > 10^{9}$ M$_{\odot}$ elliptical, spiral and irregular galaxies. \emph{Top:} pentagons show the neutral gas to stellar mass content for galaxies of all morphological types with H\,{\sc i} detections in the GASS survey \citep{CatSchKau10}. Dashed lines show the weighted median and 3\,$\sigma$ limits when non-detections are also included. \emph{Middle:} Specific star formation rate of galaxies in the KINGFISH sample \citep{SkiEngDal11}, with different symbols corresponding to different morphological types. \emph{Bottom}: same as for the middle plot, but for the dust to stellar mass ratio. In all three panels it is obvious that NGC\,4594 is a clear outlier of the relations followed by spirals, but not so if the stellar disc is considered as an independent entity (solid symbols).}
         \label{sombrero_sf_dust_smass}
   \end{figure}

It is also pertinent to ask ourselves whether the embedded disc (see Fig.\,1) displays peculiar properties.
\citet{CatSchKau10} show that the neutral gas mass fraction in the GASS sample strongly decreases as a function of stellar mass, with M$_{\star} \approx 2\times10^{11}$ M$_{\odot}$  galaxies having a weighted median M$_{HI}/$M$_{\star} = 0.016$, with a 0.386 dex scatter (see Fig.\,\ref{sombrero_sf_dust_smass}, top panel). NGC\,4594, with a stellar mass comparable to this value but a much lower neutral gas content (M$_{HI} \approx 3.1\times10^{8}$ M$_{\odot}$; \citealt{BajvanFab84}\,\footnote{All relevant quantities refer to our adopted $D=9.1$ Mpc distance.}) deviates $\approx$\,5\,$\sigma$ from this relation.
Considering instead the $5\times10^{10}$ M$_{\odot}$ disc mass would bring this ratio to within $\approx$\,2\,$\sigma$ of the observed relation in the GASS sample.
As discussed by \citet{HauBowKil08}, this massive stellar disc implies that its peak SFR must have been a factor 5--10 times higher in the past than the current 0.42 M$_{\odot}$\,yr$^{-1}$ \citep{SkiEngDal11}, and the halt of substantial disc growth is suggestive of an advanced evolutionary stage. This is supported by the analysis of stellar populations in the outer disc, that indicates old ages ($>$\,8 Gyr) and close to solar metallicities -- similar to those of the inner spheroid \citep{emsbacmon96}.
Figure \ref{sombrero_sf_dust_smass} (middle panel) shows the relation between specific SFR and stellar mass for different galaxy types in the KINGFISH sample \citep{SkiEngDal11}. NGC\,4594 again stands out as a clear outlier within spiral galaxies, but has rather normal values for ellipticals. 
Perhaps not surprisingly, the stellar disc occupies the same region as the other spirals when considered an independent entity, and the same is true for the dust-to-stellar mass content (Fig.\,\ref{sombrero_sf_dust_smass}, bottom panel).
This suggests that Sombrero's disc is not intrinsically peculiar but, instead, it is the out-of-proportion spheroid what makes the galaxy an outlier in these scaling relations.

\subsection{A scenario for Sombrero}

The emerging picture of Sombrero supports a scenario where most of the formation activity took place at an early epoch. 
The bulk of stellar mass settled in a massive and extended spheroid at $z > 2$. This spheroid exhibits a mild metallicity gradient, from roughly solar in the inner regions to a peak [Fe/H] $\approx -0.5$  in the outskirts -- the same values as for the metal-rich GCS. This latter value can be reproduced, in chemical evolution models that include gas infall, with an initial burst of gas accretion lasting $<0.2$ Gyr \citep{MouSpi10}.
Disc formation followed quickly, leaving ample time for secular evolution to take place. Dynamical resonances due to the formation of a bar could have shaped the complex structures revealed in the residual images previously shown, potentially contributing to the spheroid structure in the inner regions \citep{emsbacmon96}.

Several formation modes can in principle reproduce the observed properties, but the old inferred stellar population ages restrict them to the first 4 Gyr. Unfortunately, the differences between plausible processes (e.g. major merging, satellite accretion or dissipational formation) blur at sufficiently high redshift -- when gas absolutely dominates the baryonic content of (proto)galaxies -- making the details of the formation process hard to identify.

We can however put some constraints on the relevance of these main processes. For instance, we just pointed out that the major merger scenario fails to reproduce the extraordinarily rich GCS of NGC\,4594 -- unless progenitors were fundamentally different from any surviving disc galaxy with measured GC specific frequency. 
Low-mass satellite accretion cannot explain either the high metallicity of the outer spheroid or the extraordinarily populous metal-rich GCS. 
Both properties are however nicely consistent with a picture where cold gas flows dominate the first formation stages \citep{OseOstNaa10}. A two-stage dissipational collapse scenario (e.g., \citealt{ForBroGri97,HarHar02}) naturally results in an almost coeval, bimodal GCS with old ages and an MDF for the spheroid outer regions matching that of the metal-rich GC sub-population. Furthermore, recent hydrodynamical simulations suggest that this growth mode is prevalent for massive galaxies inhabiting low-density environments \citep{OseOstNaa10}.
The formation of the embedded disc remains a puzzle, but whatever the detailed mechanism, it requires the infall of an enormous amount of high-angular momentum pre-enriched gas in order to reproduce the solar and super-solar metallicities of the outer and inner rings \citep{emsbacmon96}.

Finally, we stress that our study suggests that what has been called `bulge' in NGC 4594 is likely not a classical bulge. We show indications that it is either a multi-component spheroid, or a system resembling regular intermediate-mass elliptical galaxies. 
Ultra deep, multi-wavelength imaging of Sombrero could reveal new clues about its extended spheroidal component.

\section*{Acknowledgments}
We are grateful to Alister Graham, Peter Erwin, and an anonymous referee for helpful comments. It is a pleasure to thank Tom Jarrett for producing the S$^4$G PSF image. This research has made use of NASA's Astrophysics Data System and the NASA/IPAC Extragalactic Database (NED), which is operated by the Jet Propulsion Laboratory, California Institute of Technology, under contract with the National Aeronautics and Space Administration. We acknowledge the usage of the HyperLeda database (http://leda.univ-lyon1.fr).

\bibliographystyle{mn2e}
\bibliography{../gadotti_refs}

\appendix
\section{Fitting the rings}
\label{app:rings}

\begin{figure}
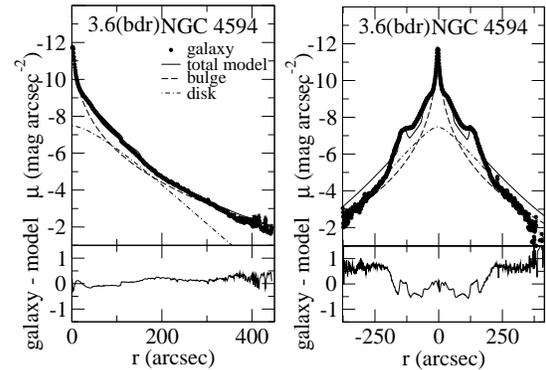

   \centering
   \includegraphics[keepaspectratio=true,width=3.5cm,clip=true]{N4594_bdr.eps}
   \includegraphics[keepaspectratio=true,width=3.5cm,clip=true]{N4594vec_bdr.eps}
   \caption{Left: surface brightness radial profiles (in arbitrary units) obtained through ellipse fits to the IRAC image and to the {\sc galfit} model images of a decomposition including bulge, disc and two rings. Right: the same radial profiles, but obtained through a cut along the disc major axis. Lower panels show residual profiles after subtracting the full model profile from the galaxy profile.}
   \label{fig:profs_bdr}
\end{figure}

\begin{figure}
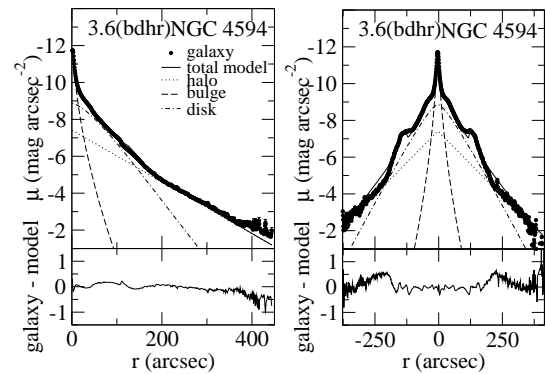

   \centering
   \includegraphics[keepaspectratio=true,width=3.5cm,clip=true]{N4594_bdhr.eps}
   \includegraphics[keepaspectratio=true,width=3.5cm,clip=true]{N4594vec_bdhr.eps}
   \caption{Same as Fig. \ref{fig:profs_bdr}, but for a decomposition including bulge, disc, halo and two rings.}
   \label{fig:profs_bdhr}
\end{figure}

\begin{figure*}
   \centering
   \includegraphics[keepaspectratio=true,width=5cm,clip=true]{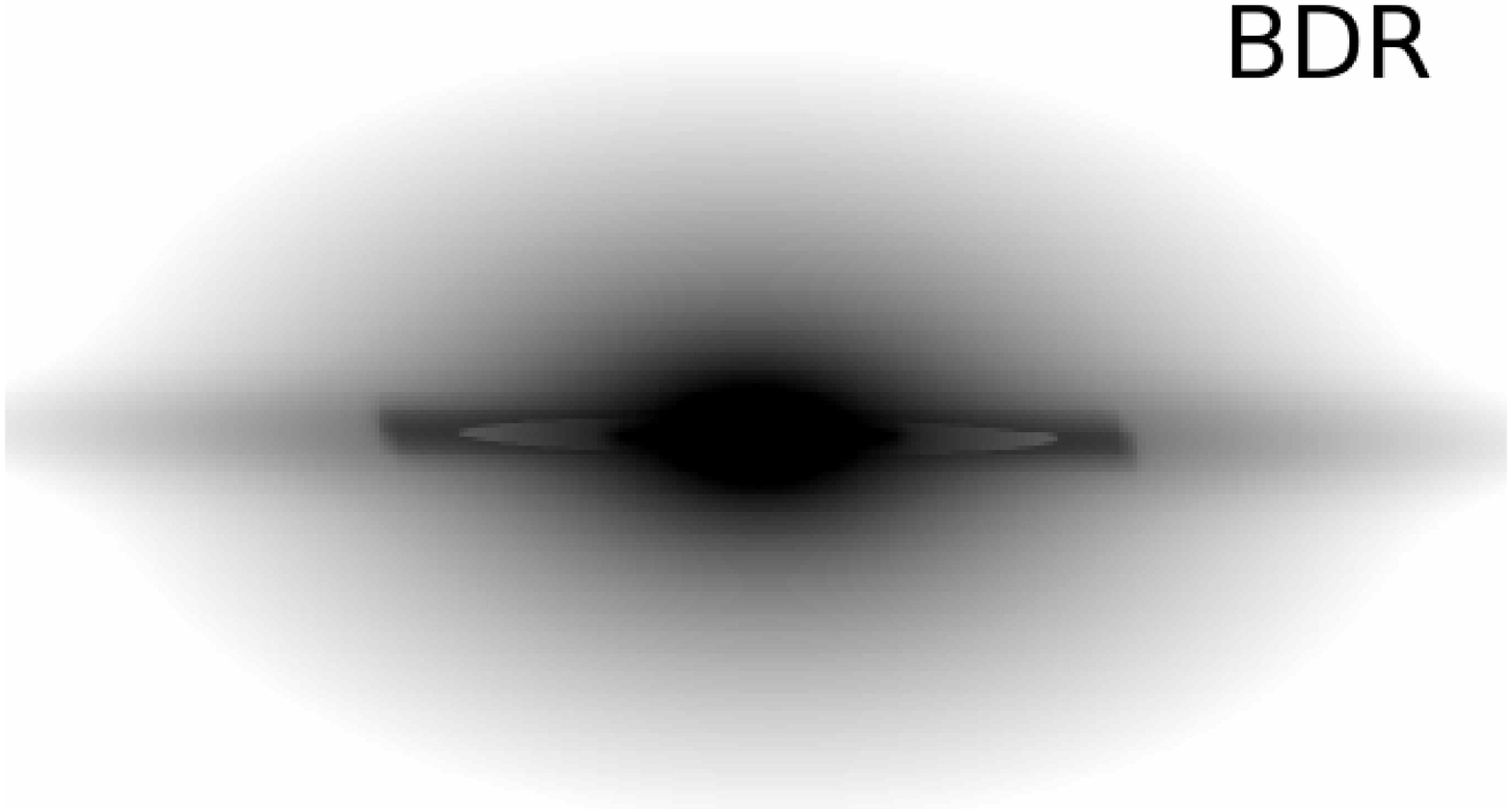}
   \includegraphics[keepaspectratio=true,width=5cm,clip=true]{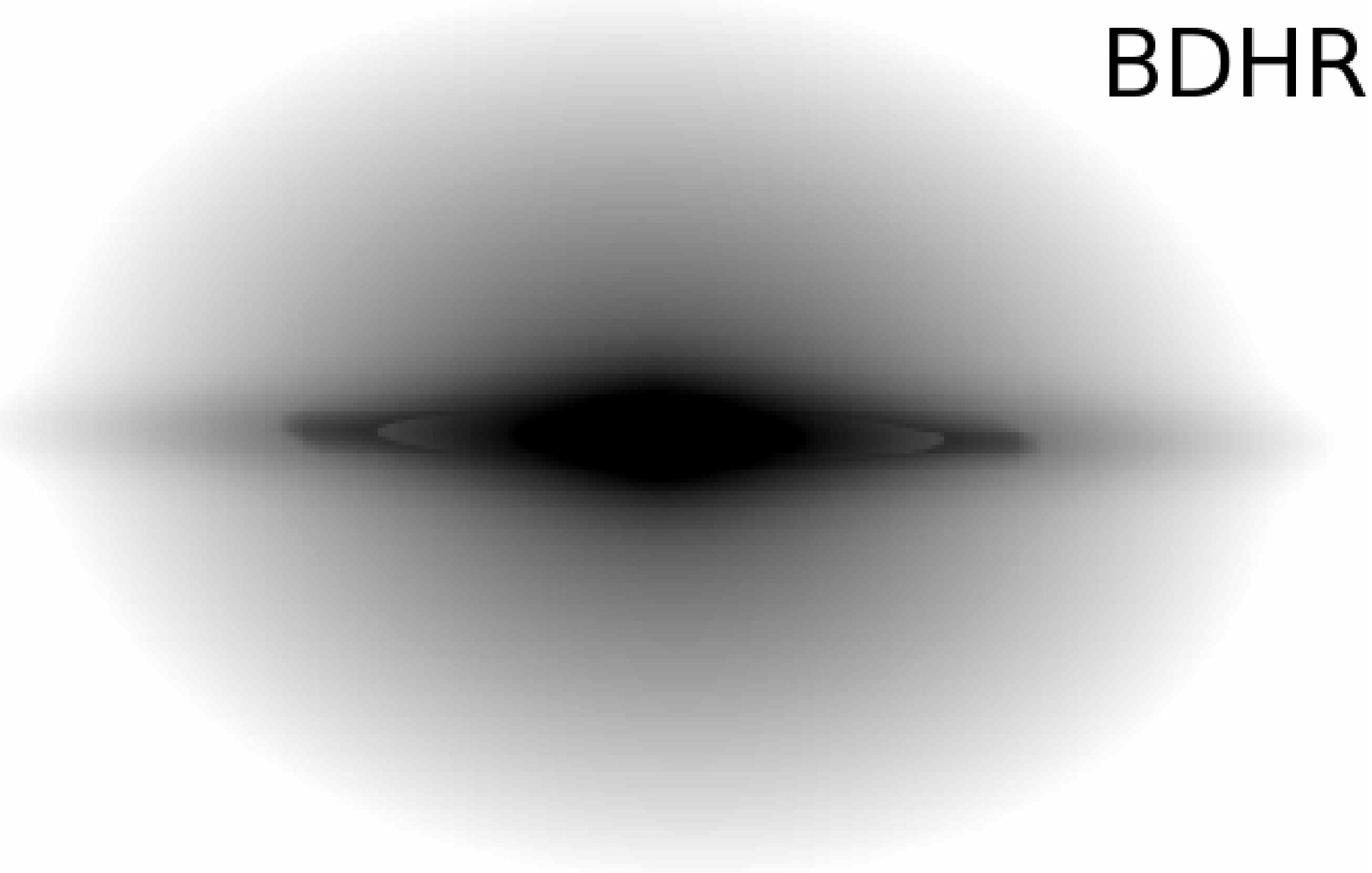}\\
   \includegraphics[keepaspectratio=true,width=5cm,clip=true]{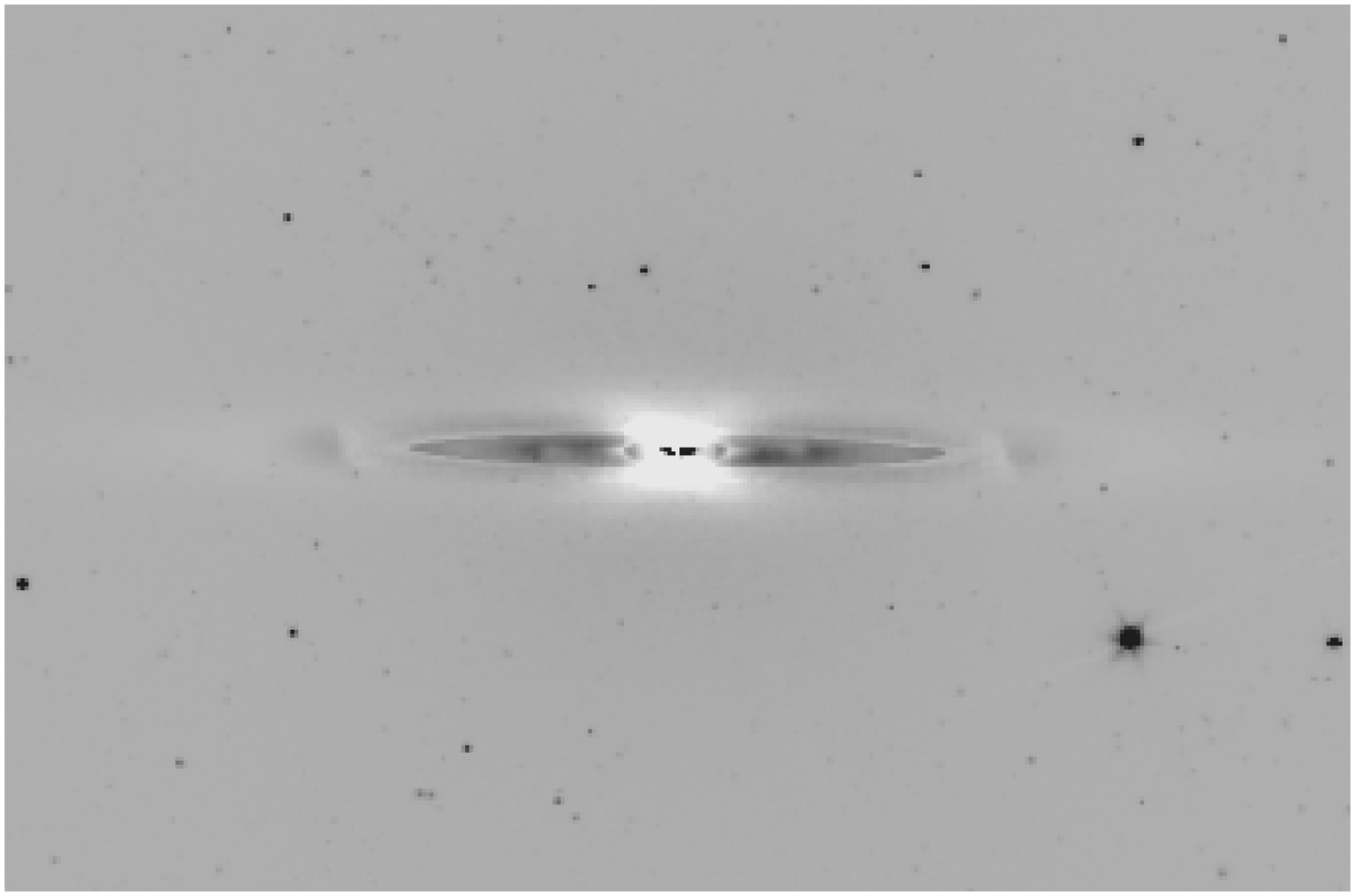}
   \includegraphics[keepaspectratio=true,width=5cm,clip=true]{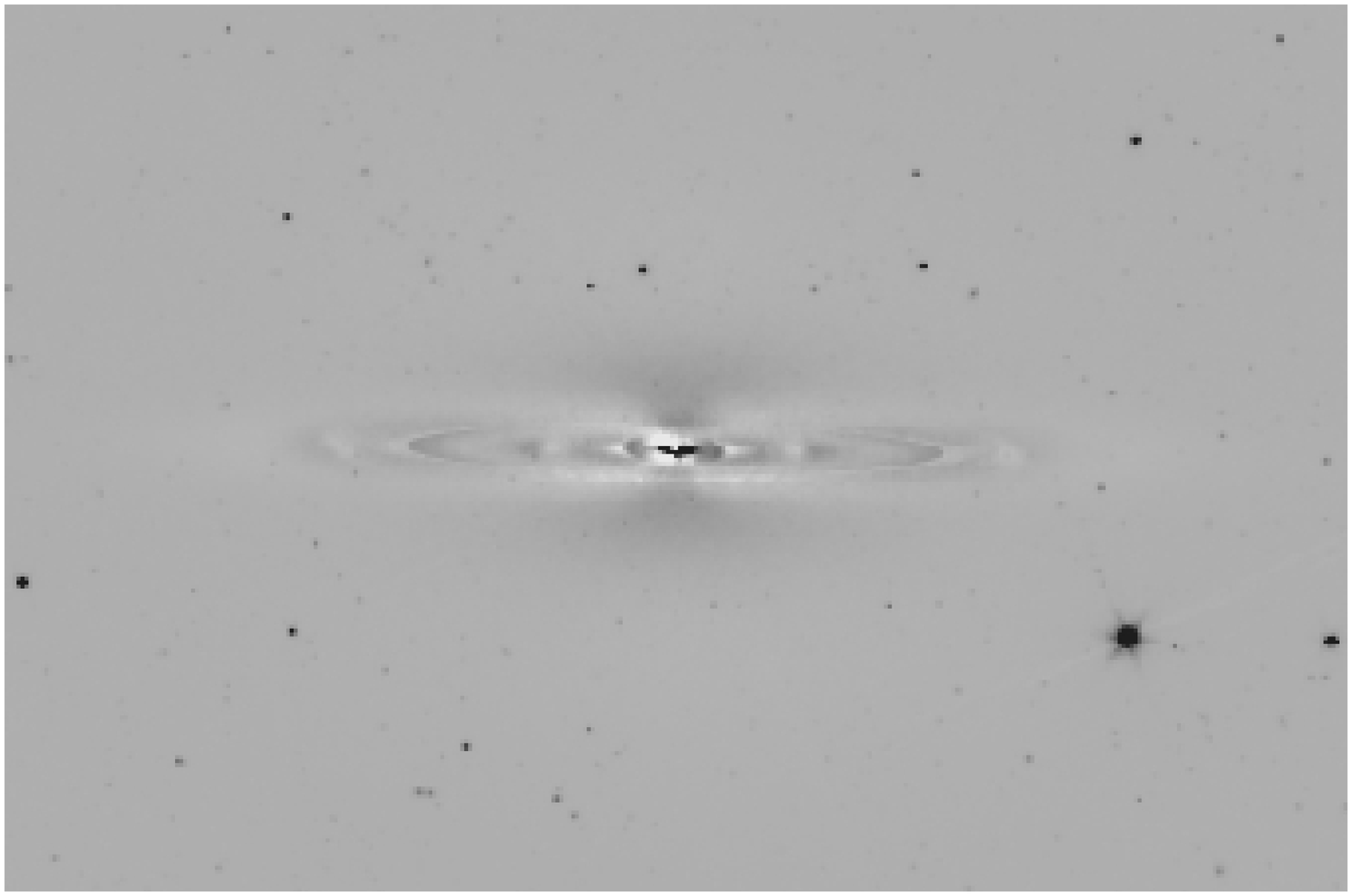}
   \caption{Images of the full {\sc galfit} models (top) and the corresponding residual images (bottom) after subtraction of the models from the IRAC image. Left: model {\sc bdr} (bulge+disc+rings); right: model {\sc bdhr} (bulge+disc+halo+rings). The images are in the same scale and display levels as in Fig. \ref{fig:imgs}.}
   \label{fig:imgs_r}
\end{figure*}

With the aim of testing whether the results and fits performed in Sect. \ref{sec:struc} are affected by the fact that such fits do not include components to model the two main stellar rings in Sombrero, we used the latest version of {\sc galfit} \citep{PenHoImp10}, which allows one to model structures such as rings. The rings are described as elliptical Gaussian functions with inner and outer truncations. We first tried to start the fits from scratch. However, the parameters describing the rings, if left free, result in model components which are essentially merged with the galaxy disc, thus failing to reproduced the bumps observed in Sombrero's surface brightness radial profiles, in particular along the disc major axis. We thus had to fix all parameters describing the rings, after many manual interactions. For this reason, and because the resulting models for bulge, disc and halo do not change significantly after including the rings, we keep using models {\sc bd} and {\sc bdh} from the analysis above as reference models.

The results from these new fits are shown in Figs. \ref{fig:profs_bdr} to \ref{fig:imgs_r}. The new model with bulge and disc plus rings is called {\sc bdr}, while the corresponding model with an extra halo component is called {\sc bdhr}. Figure \ref{fig:profs_bdr} should be compared with Fig. \ref{fig:profs_bd}, while Fig. \ref{fig:profs_bdhr} should be compared with Fig. \ref{fig:profs_bdh}. As mentioned, the models for bulge, disc and halo do not change significantly after adding the ring components. The variations observed at each structural parameter are within the uncertainties found with {\sc budda} for models {\sc bd} and {\sc bdh}, which are typically $\approx10\%$.

Figure \ref{fig:imgs_r} should be compared with Fig. \ref{fig:imgs}. One sees that after the inclusion of the rings it is still evident that the extra halo component results in an improved fit.

\label{lastpage}

\end{document}